\documentclass[reprint,superscriptaddress,amssymb,amsmath,sn,prxq]{revtex4-2}

\usepackage{floatrow}
\usepackage{graphicx}
\usepackage{amssymb}
\usepackage{epstopdf}
\usepackage{upgreek}
\usepackage{dcolumn}
\usepackage{bm}
\usepackage{gensymb}
\usepackage{braket}
\usepackage{amsmath}
\usepackage{mathtools}
\usepackage{float}
\usepackage{tikz}
\usepackage[colorlinks=true,allcolors=blue]{hyperref}

\usepackage{array}
\newcolumntype{C}[1]{>{\centering\arraybackslash}m{#1}}

\usepackage{xcolor}

\usepackage[resetlabels,labeled]{multibib}
\usepackage{xfrac}

\usepackage{comment}

\usepackage{hhline}

\usepackage[showframe=false]{geometry}
\usepackage{enumitem}

\usepackage{amsmath}

\usepackage{physics}

\usepackage{lmodern}
\usepackage{microtype}

\allowdisplaybreaks

\expandafter\ifx\csname package@font\endcsname\relax\else
\expandafter\expandafter
\expandafter\usepackage
\expandafter\expandafter
\expandafter{\csname package@font\endcsname}%
\fi

\newcommand{\micro}{$\upmu$}

\newcommand{\be}{\begin{eqnarray}}
\newcommand{\ee}{\end{eqnarray}}
\newcommand{\bfig}{\begin{figure}}
	\newcommand{\efig}{\end{figure}}

\definecolor{myorange}{rgb}{0.97,0.59,0.27}

\DeclareFontFamily{U}{mathb}{}
\DeclareFontShape{U}{mathb}{m}{n}{
	<-5.5> mathb5
	<5.5-6.5> mathb6
	<6.5-7.5> mathb7
	<7.5-8.5> mathb8
	<8.5-9.5> mathb9
	<9.5-11.5> mathb10
	<11.5-> mathbb12
}{}

\makeatletter
\@ifundefined{selectlanguage}{\newcommand{\selectlanguage}[1]{}}{\renewcommand{\selectlanguage}[1]{}}
\makeatother

\begin{document}
\makeatletter
  \@namedef{figure}{\killfloatstyle\def\@captype{figure}\FR@redefs
    \flrow@setlist{{figure}}%
    \columnwidth\columnwidth\edef\FBB@wd{\the\columnwidth}%
    \FRifFBOX\@@setframe\relax\@@FStrue\@float{figure}}%
\makeatother

\setlist[itemize]{wide = 0pt}

\title{A Traveling-Wave Parametric Amplifier With Integrated Diplexers}% 

\author{C. Denney}
\affiliation{National Institute of Standards and Technology, 325 Broadway, Boulder, CO 80305, USA}
\affiliation{Colorado School of Mines, 1500 Illinois St, Golden, CO, 80401, USA}
\author{K. Genter}
\affiliation{National Institute of Standards and Technology, 325 Broadway, Boulder, CO 80305, USA}
\affiliation{University of Colorado, 2000 Colorado Ave., Boulder, CO 80309, USA}
\author{K. Cicak}
\affiliation{National Institute of Standards and Technology, 325 Broadway, Boulder, CO 80305, USA}
\author{J. D. Teufel}
\affiliation{National Institute of Standards and Technology, 325 Broadway, Boulder, CO 80305, USA}
\affiliation{University of Colorado, 2000 Colorado Ave., Boulder, CO 80309, USA}
\author{J. Aumentado}
\affiliation{National Institute of Standards and Technology, 325 Broadway, Boulder, CO 80305, USA}
\author{F. Lecocq}
\email{florent.lecocq@nist.gov}
\affiliation{National Institute of Standards and Technology, 325 Broadway, Boulder, CO 80305, USA}
\affiliation{University of Colorado, 2000 Colorado Ave., Boulder, CO 80309, USA}
\author{M. Malnou}
\email{maxime.malnou@nist.gov}
\affiliation{National Institute of Standards and Technology, 325 Broadway, Boulder, CO 80305, USA}
\affiliation{University of Colorado, 2000 Colorado Ave., Boulder, CO 80309, USA}

%, K. Genter, K. Cicak, J. D. Teufel, J. Aumentado, F. Lecocq, M. Malnou

\date{\today}% It is always \today, today,

\begin{abstract}
Traveling-Wave Parametric Amplifiers (TWPAs) are ubiquitous in superconducting circuit readout, providing high gain with near-quantum-limited noise performance across a wide bandwidth. Their operation, however, relies on a strong microwave pump tone that is typically delivered using off-chip passive components, such as directional couplers or diplexers. These external elements add loss, increase system complexity, and limit scalability. Here, we present a traveling-wave parametric amplifier that incorporates on-chip input and output diplexers for pump routing. This co-fabricated architecture offers a compact and scalable solution for superconducting-circuit readout.
\end{abstract}

\maketitle

\section{Introduction}
% Let's just write the argumentation of why we did this work. What's the motivation.
Parametric amplifiers have become indispensable for measuring signals interacting with superconducting circuits, including the microwave resonators used in qubit and sensor architectures \cite{Abdo2014Josephson,brubaker2017first,du2018search,heinsoo_rapid_2018,arute2019quantum,Lecocq2021efficient,Rosenthal2021efficient,Backes2021a,Krinner2022realizing,malnou2023improved}. Often used as the first stage of amplification, they provide high gain with minimal added noise \cite{macklin_nearquantum-limited_2015,planat2020photonic,malnou_three-wave_2021,malnou_low-noise_2024}. Among them, traveling-wave parametric amplifiers (TWPAs) have become increasingly popular over the past decade. Compared to resonant parametric amplifiers, the TWPA's inherently broader bandwidth and higher power handling allows them to read out multiple resonators simultaneously, greatly improving practicality in multiplexed readout architectures \cite{Krinner2022realizing, malnou2023improved}.
%Their inherently broadband response eliminates the need to tune the amplifier to individual resonator frequencies and their higher power handling allows for multiple resonators to be probed simultaneously, greatly improving practicality in multiplexed readout architectures compared to resonant parametric amplifiers \cite{aumentado_superconducting_2020}.

Yet, existing TWPAs still rely on a substantial amount of external microwave hardware that impede scalability\cite{macklin_nearquantum-limited_2015,planat2020photonic,malnou_three-wave_2021,Krinner2022realizing,malnou2023improved,malnou_low-noise_2024,malnou_travelling-wave_2025,Ranadive2025a}. In conventional implementations, the amplifier chip must be surrounded by off-chip components that route and manage the strong pump tone required for parametric gain. Reciprocal elements such as directional couplers or diplexers are used to inject the pump into the signal path and to filter it out at the output. In addition, nonreciprocal devices such as isolators are typically placed between the TWPA and the device under test to suppress back-action. These input/output (I/O) networks, assembled from connectorized commercial components, degrade the readout chain's noise performance and increase the overall system footprint.

In this article, we present a four-wave-mixing Josephson TWPA with superconducting diplexers co-fabricated on chip at both its input and output. These diplexers introduce two additional physical ports that enable local pump delivery and pump extraction without external microwave hardware. They also separate the signal and idler bands into distinct off-chip paths, eliminating the requirement for the upstream readout chain to also be matched at the idler frequency. This diplexed TWPA (D-TWPA) delivers about $13$\,dB of gain across a $2$\,GHz signal-amplification bandwidth. Within this band and for this gain, the average system-added noise is $2$ quanta, as determined from a noise measurement employing a shot-noise tunnel junction \cite{Spietz2003primary,spietz2006shot,chang2016noise,malnou_three-wave_2021,malnou_low-noise_2024}.

%simplifying the requirements for the upstream readout chain to present a good impedance match at the idler frequency
\section{Design}
\begin{figure*}[htbp!]
    \centering
    \includegraphics[width=1\linewidth]{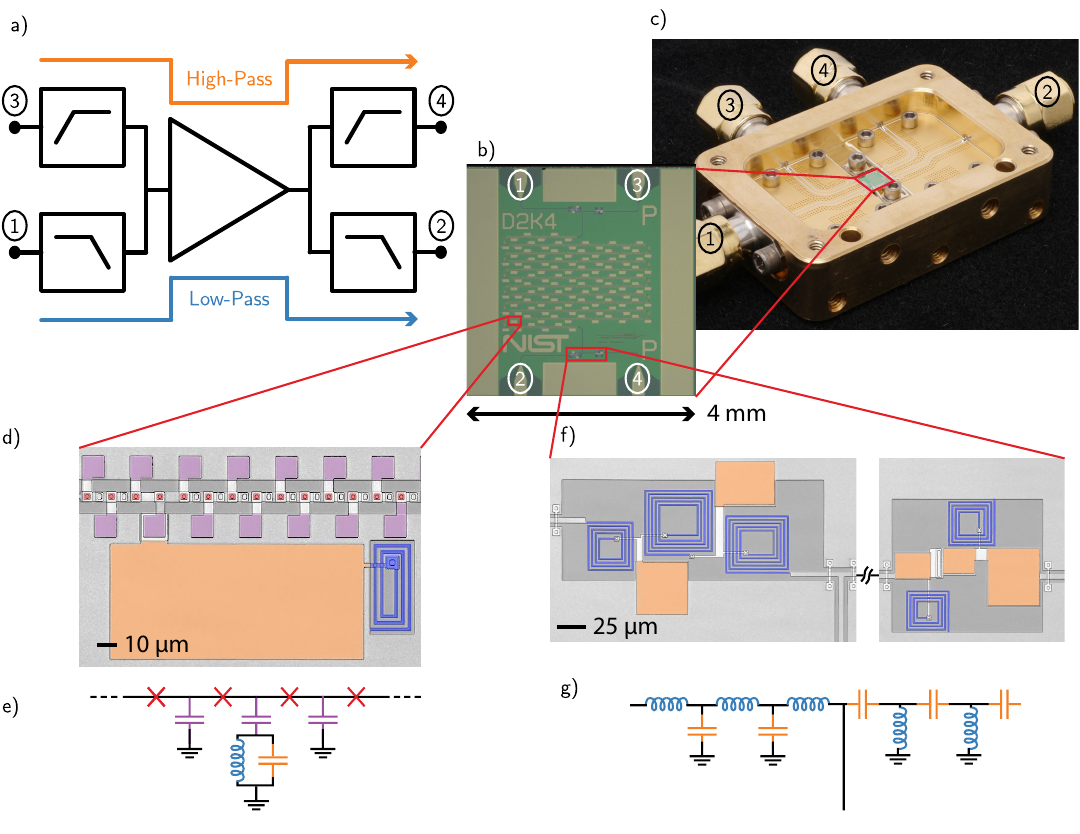}
    \caption{An overview of the D-TWPA integrated layout. (a) The TWPA is connected to the common port of a diplexer on each end. (b) The resulting chip has four microwave ports, along with the packaged device (c). The phase-matching resonator (d) and the lumped-element diplexers (f) are shown in false color micrographs, along with their corresponding equivalent schematics, (e) and (g) respectively.}
    \label{fig:concept}
\end{figure*}

% 1 paragraph describing the TWPA. JJ, rpm.
The active part of the D-TWPA consists of a standard 50\,\ohm{} nonlinear artificial transmission line formed by series Josephson-junction inductors shunted to ground through capacitors, see Fig.\,\ref{fig:concept}b. This amplifier employs four-wave mixing (4WM) where the pump frequency $\omega_p$ lies between the signal ($\omega_s$) and idler ($\omega_i$) frequencies, satisfying $2\omega_p=\omega_s+\omega_i$ \cite{yaakobi_parametric_2013}. To achieve phase matching between the pump, signal and idler propagating waves, an LC resonator is periodically inserted in the shunt branch \cite{obrien_resonant_2014}, see Fig.\,\ref{fig:concept}d,e. The resonator is designed with a resonance at $8.7$\,GHz, slightly above the desired pump frequency at $8.5$\,GHz, and its low mode impedance ($Z_\mathrm{rpm}\approx15$\,\ohm{}) minimizes the width of the generated stopband, which is unusable for amplification.

% First, talk about what we're gonna separate: signal band on one side, and pump + idler band on the other side
For practical operation, we aim to separate the low-frequency signal band from the higher-frequency band that contains both the pump and idler tones. We therefore designed a pair of identical lumped-element diplexers whose common ports connect directly to the TWPA, see Fig.\,\ref{fig:concept}a. This configuration creates a four-port device (two low-frequency ports and two high-frequency ports), as shown in Fig.\,\ref{fig:concept}c. Each diplexer is constructed from fifth-order Chebyshev low-pass and high-pass filters (see Fig.\,\ref{fig:concept}g). Their compact footprint (approximately $300$\,\micro m\,$\times$\,$300$\,\micro m, see Fig.\,\ref{fig:concept}f), much smaller than the TWPA itself, allows them to be readily integrated on chip without significantly increasing the chip area. 

% Values in appendix ?

%figure comments: Maybe get rid of green, make 3 and 4 straight, Sans Serif font, make the HP / LP colors distinct from the false colors, label the common port / LP / HP ports, add straight line on leftmost inductor, shrink the schematics, cut down the x axis on the RPM image, maybe arrow from triangle to a and DP to b

%(a) The JJ-based artificial transmission line is loaded every 20 unit cells with a capacitively coupled tank circuit. (b) Each side of the TWPA is equipped with a high-pass/low-pass diplexer constructed from thin-film lumped-element components. (c) The final device has four physical ports, two for the diplexers on each side of the TWPA. Panels (d), (e) and (f) are optical microscope images showing the D-TWPA components described by their electrical schematics in (a), (b) and (c) respectively.

The diplexer crossover frequency is chosen so that the signal band lies below $8$\,GHz, ensuring compatibility with a standard C-band readout chain. Conversely, the TWPA idler-band output is routed to a dedicated port that can be terminated in a cryogenic matched load mounted on the TWPA package, eliminating the need for the rest of the chain to provide a good match at these higher frequencies. The common ports of the diplexers are well matched across the entire operating band.

The TWPA and diplexers were co-fabricated on a single chip using niobium trilayer Josephson junctions (critical current $I_c\simeq5$\,\micro A), and parallel-plate capacitors employing a low-loss amorphous silicon dielectric ($\tan\delta\simeq4\times10^{-4}$) \cite{lecocq2017nonreciprocal,malnou_travelling-wave_2025}.

\section{Scattering Measurements}

% Include method, simulation details, unpmped case, ...
% " We placed the device in a DR with some switches " Blah blah blah
%
% Mention the thrus in the caption, main tex. Then a dashed line for the thru reference

% We measured S21 using a thru reference only.
% With the pump off, we can see that the insertion loss is smaller than some other devices because of our low-loss process.
% With the pump off we can see the diplexer crossover is near 8.5 GHz and the phase matching feature puts the pump near 9 GHz.

% With the pump on we get about 15 dB of gain when the pump is optimized for added noise. The gain is at the level from 6-8 GHz which aligns with the readout frequency of common transmon qubit processors.

%Transmission measurements between the signal ports and idler ports of the device are shown in Fig. \ref{fig:scattering} (a). These measurements were calibrated against an SMA thru, making the assumption that the networks between the VNA and the DUT are well-matched on both ends. The diplexer crossover occurs near 8.5 GHz at almost exactly -6 dB, the ideal case for two diplexers back-to-back. The dispersion feature appears on the idler transmission measurement around 9 GHz. The insertion loss at 14 GHz is 4 dB.

%In Fig. \ref{fig:scattering} (b) you can see the transmission with the pump on. This is the operating point that optimized the system added noise. 

We first characterized the device's driven response at cryogenic temperatures. Microwave switches mounted on each of the four ports allowed us to compare the TWPA's transmission to a through reference in both the signal and idler bands (see  the full characterization setup in appendix \ref{app:setup}).
% Calibration of this kind assumes that both the device and the networks connecting the device to the VNA are well-matched. We estimate that the match of these networks would be no worse than -10 dB, which corresponds to a transmission error of less than 1 dB in the worst case. \con{Do we need to elaborate? Does this deserve an equation in the main text? A mention in an appendix?}

\begin{figure}[H]
    \centering
    \includegraphics[width=1\linewidth]{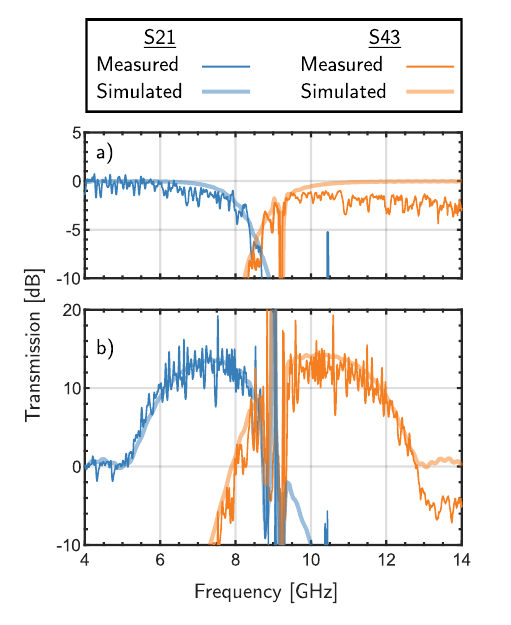}
    \caption{Gain and insertion loss measurements relative to a through reference. With the pump off (a) the crossover of the diplexer appears at $8.5$\,GHz, and the resonant phase matching feature is just above $9$\,GHz. (b) The gain profile with a pump optimized for noise performance ($9.06$\,GHz and $-76.8$\,dBm) is measured through both $S_{21}$ (blue curve) and $S_{43}$ (orange curve). Predicted scattering from simulations neglecting internal loss and early junction switching are show in light blue and orange, in good agreeement with the measurements.}
    %NOTES: If we use experimental parameters we get this simulation curve
    \label{fig:scattering}
\end{figure}

Figure \ref{fig:scattering}a shows the transmission amplitude through the low- and high-frequency ports of the unpumped TWPA, referenced to the through-cable calibration. The thickness of the dielectric layer was measured to be 13\,\% thicker than the value used for simulation, which caused the diplexer crossover and the resonant phase matching feature to shift to $8.5$\,GHz and $9.2$\,GHz respectively. The insertion loss remains below $3$\,dB up to $14$\,GHz, owing to the low-loss amorphous-silicon dielectric.

%The $6$\,dB diplexer crossover appears near $8.5$\,GHz, slightly higher than the design target due to a slight difference in the actual dielectric thickness from the target. In the idler band, the resonant phase matching feature produces a sharp, narrow dip in transmission just above $9$\,GHz, defining the upper edge of the usable pump frequency.

\begin{figure*}[htbp!]
    \centering
    \includegraphics[width=1\linewidth]{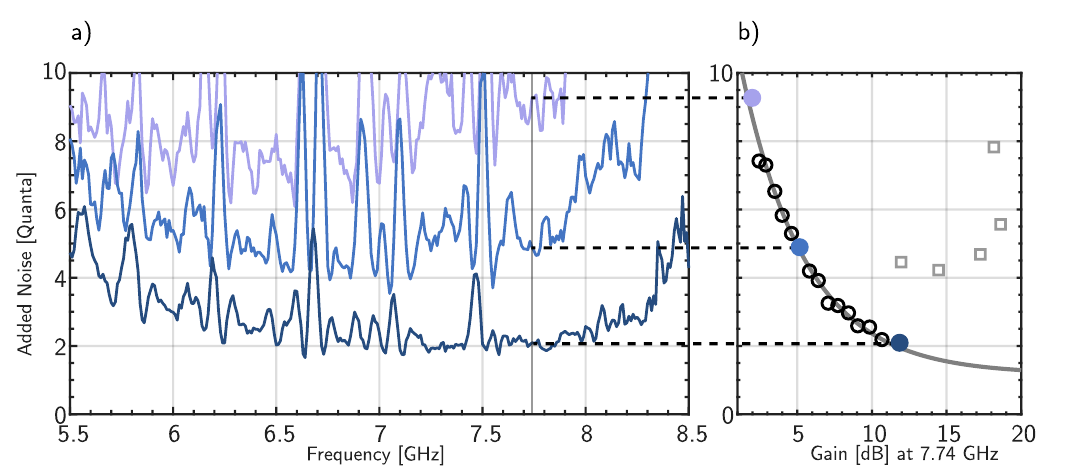}
    \caption{Measurement of the chain-added noise where the TWPA is the first amplifier. (a) The chain-added noise between $5.5$\,GHz and $8.5$\,GHz is shown for three different TWPA gains increasing from light purple to dark blue.. (b) Focusing on a frequency bin centered at $7.74$\,GHz, the chain-added noise decreases as the TWPA gain increases, until it increases again at higher pump powers These measurements (squares) were excluded from the fit used to extract $N_\mathrm{TWPA}$.}
    \label{fig:noise}
\end{figure*}

Figure \ref{fig:scattering}b shows the calibrated transmission amplitude through the low- and high-frequency ports of the device ($S_{21}$ and $S_{43}$ respectively) when the pump is applied through the high-frequency input port. The TWPA generates broadband gain of about $13$\,dB between $6$ and $8$\,GHz in the signal band, and between $10$ and $12$\,GHz in the idler band. We attribute the $\sim5$\,dB gain ripples to impedance mismatches that arise between the TWPA chip and its packaging when it is pumped (see appendix \ref{app:packaging}). Although a maximum broadband gain of $\sim20$\,dB can be obtained at higher pump power, the noise performance degrades at these higher gains \cite{peng_floquet-mode_2022,wang2025high}.

% The phase matching feature and the diplexer crossover moved in common-mode across the wafer which preserves the ability for the pump to be coupled through the idler port of the diplexers. Ultimately, this device is fit for deployment in frequency-multiplexed qubit readout architectures with 6 - 8 GHz readout resonator frequencies.

% We simulated this TWPA by modeling the RPM cells in a EM solver, then importing this model into wrSPICE with the JJs and caps. This simulation uses a pump power derived from experiment.

We compared these transmission measurements to the steady-state response of an equivalent lumped-element model of the TWPA and diplexers, simulated in the time domain using WRSpice \cite{wrspice}. The pump power of $-76.8$\,dBm used in the simulation was chosen to match our estimate of the pump power incident on the TWPA’s high-frequency input in the experiment (see appendix \ref{app:powercal}). As shown in Fig.\,\ref{fig:scattering}, this model accurately reproduces both the unpumped (a) and pumped (b) transmission responses, except for internal losses which were not modeled.

% Simulation of the device gain is also pictured in Fig. \ref{fig:scattering} (b). After adjusting our simulations to account for a fabrication variation in the dielectric thickness and the experimental pump power, S-Parameters of the resonant phase matching cell were converted into a SPICE subcircuit and populated in a lumped-element approximation of the TWPA using the built-in JJ and capacitor models from wrSPICE. A transient analysis was performed until a steady state was reached, and the Fourier transforms of the input and output waveforms were used to calculate the S21. Dielectric loss in the capacitors was not simulated.

\section{Noise Characterization}

We extracted the broadband added noise of the chain using a Y-factor measurement, with a shot-noise tunnel junction (SNTJ) serving as a calibrated noise source \cite{Spietz2003primary,spietz2006shot,chang2016noise,malnou_three-wave_2021,malnou_low-noise_2024}. The microwave switch connected to the TWPA’s low-frequency input can toggle between the continuous-wave probe tone and the SNTJ, allowing us to perform scattering and noise measurements under identical conditions \textit{in situ}. The output noise $N_\mathrm{out}$ was recorded on a spectrum analyzer as a function of the chain's input noise $N_\mathrm{in}$ generated by the SNTJ.

To extract the chain-added noise $N_\mathrm{add}$ (in units of quanta) we fit the measured $N_\mathrm{out}$ as a function of $N_\mathrm{in}$ using \cite{malnou_low-noise_2024}
\begin{equation}
    N_\mathrm{out} = G_\mathrm{tot}(N_\mathrm{in} + N_\mathrm{add})
    \label{eq:AddedNoise}
\end{equation}
Where $G_\mathrm{tot}$, the gain of the readout chain, is a fit parameter. The added noise includes the low-gain correction
\begin{equation}
    N_\mathrm{add} = \frac{G_\mathrm{TWPA}-1}{2G_\mathrm{TWPA}}+N_\mathrm{ex},
\end{equation}
where $G_\mathrm{TWPA}$ is the gain of the TWPA alone. We assume a $14$\,mK idler temperature that contributes to $1/2$ quantum of noise since the diplexer allows the SNTJ to illuminate only the signal input (and not the idler). The term $N_\mathrm{ex}$ represents the excess of noise above the quantum limit of $N_\mathrm{add}=1/2$ attained at high gain.

Figure \ref{fig:noise}a shows the extracted $N_\mathrm{add}$ as a function of frequency for various pump powers. As the power (and gain) increases, $N_\mathrm{add}$ decreases, reaching an average minimum of $2$ quanta between $6$ and $8$\,GHz. However, once the gain exceeds roughly $13$\,dB, the added noise begins to rise again as shown in Fig.\,\ref{fig:noise}b, illustrating the gain--noise tradeoff typically seen in TWPAs \cite{wang2025high,peng_floquet-mode_2022}.

By measuring the chain added noise, $N_\mathrm{add}$, as a function of $G_\mathrm{TWPA}$ we can separate the noise contribution of the TWPA itself, $N_\mathrm{TWPA}$, from that of the remaining amplification chain, $N_\mathrm{rem}$, using
\begin{equation}
    N_\mathrm{add} = N_\mathrm{TWPA} + \frac{N_\mathrm{rem}}{G_\mathrm{TWPA}}.
\end{equation}
Fitting $N_\mathrm{add}$ as a function of $G_\mathrm{TWPA}$ (known from thru-calibrated S parameters during the same cooldown) yields $N_\mathrm{TWPA}=1.17\pm0.14$ quanta $7.74$\,GHz, see Fig.\,\ref{fig:noise}b.

% Make the fit line black. Update the schematic. Make the fonts Sans Serif. Make sure the lines are centered on the right circles. The blue is too strong compared to the mauve. Maybe try partial transparency on the two upper curves. Maybe try adjusting the connecting lines -- thinner, dashed?. Make the red less jumpy -- try crosses or squares or triangles

\section{Conclusion}
We have presented a TWPA with integrated input and output diplexers, eliminating the need for external circuitry to deliver and filter the microwave pump. This architecture improves the practicality of the amplifier and reduces its overall footprint, making it well suited for deployment in large-scale quantum computing systems. The diplexed TWPA achieves performance on par with existing devices in both gain and added noise. Looking forward, such on-chip diplexers can be combined with the various recent advances in TWPA technology \cite{gaydamachenko_rf-squid-based_2025, wang2025high, chang_josephson_2025} that have lead to higher gain and power handling, lower ripples, and reduced added noise.

\appendix

\section{Amplifier packaging}
\label{app:packaging}

% We packaged this chip in a box with 4 SMA connectors with PCB/wirebond transitions.

The amplifier is clipped directly to a brass housing (see Fig.\,\ref{fig:concept}c) alongside printed circuit-boards (PCBs) that facilitate the transition from the coaxial connector to the wirebonds. Toward the chip, single-stub matching was used to compensate the wire-bond inductance. Toward the coaxial connector, the overall width of the coplanar waveguide (CPW) on the PCB was optimized to eliminate the need for any compensation. The simulated return loss at the signal port for a perfectly $50$\,\ohm{} load at the common port of the diplexer (connected to the TWPA) was $-17.5$\,dB at $18$\,GHz. The final module has four ports, with low- and high-frequency inputs and outputs.

%The amplifier chips were packaged in a brass module and wirebonded to a printed circuit board (PCB) that faciliated the connection to the four coaxial connectors. Single-stub matching was implemented to compensate the wirebond inductance and the overall length of the coplanar waveguide (CPW) was optimized to reduce reflections at the coaxial transition. We simulated the scattering from the coaxial connector, through the PCB, and onto the on-chip CPW mode in Ansys HFSS and achieved a return loss of $17.5$\,dB at 20 GHz.

\section{Experimental Setup}
\label{app:setup}

\begin{figure*}[htbp!]
    \centering
    \includegraphics[width=1\linewidth]{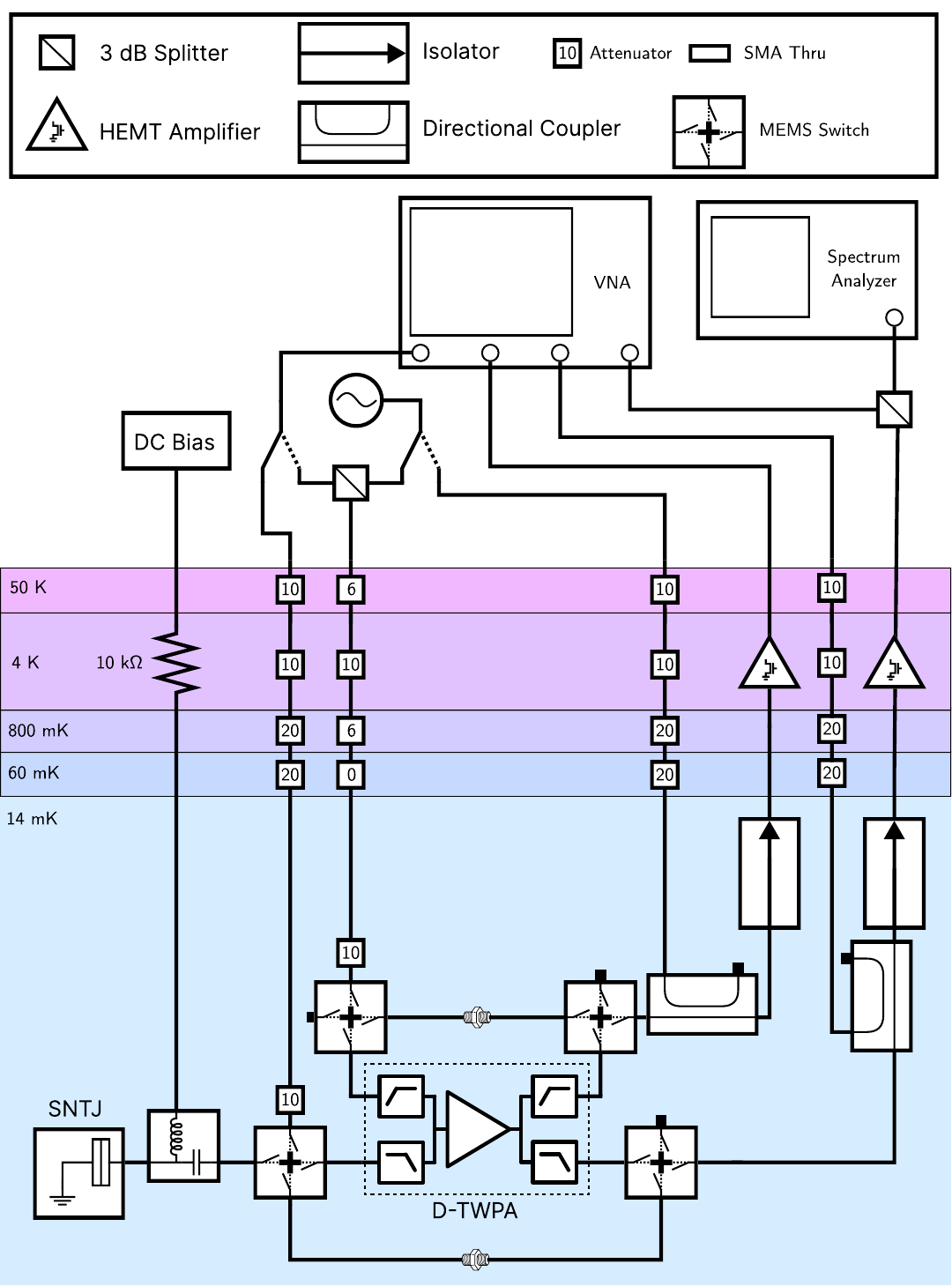}
    \caption{Full schematic of the experimental setup used to characterize the D-TWPA.}
    \label{fig:testSetup}
\end{figure*}

The TWPA was cooled to $14$\,mK with a dilution refrigerator. Each of the device's four ports was connected to one port of a 4-port micro electromechanical systems (MEMS) switch. These switches were configured to bypass the high-pass and low-pass paths through the TWPA with an SMA thru to calibrate the transmitted power through the device. The switch attached to port 1 of the TWPA (the low-pass input) was also connected to an SNTJ for \textit{in-situ} noise characterization. The TWPA outputs were connected to $20$\,dB directional couplers so that scattering measurements counter-propagating with the pump could be performed. Finally, each of the outputs was fed into an isolator followed by a high electron mobility transistor (HEMT) amplifier at $4$\,K. The full testing setup is shown in Fig. \ref{fig:testSetup}.

\section{Cryogenic Power Calibration}
\label{app:powercal}

% In room-temperature measurements, a simple 2-port VNA measurement can be taken of the test fixture to determine the insertion loss, then the known power from the VNA can be used to calculate the power delivered to the device. However, at cryogenic temperatures we do not have access to both ports of the fixture when it is cold, and the insertion loss changes significantly between the room-temperature and cryogenic states. Therefore, we must use a different method to calibrate the power delivered to the reference plane of the TWPA.
Knowledge of the power incident on the ports of the TWPA is critical for properly characterizing the $1$\,dB compression point and the actual pump power. We used the known power spectral density (PSD) $S_\mathrm{in}$ from the SNTJ to calibrate this power. First, we configured the cryogenic switches to connect the SNTJ directly to the readout chain, bypassing the TWPA using the SMA thru. We measured the output PSD $S_\mathrm{out}$ at room-temperature for a range of SNTJ biases. It is a function of the known input PSD, $S_\mathrm{in}$, and the gain $G$ of the readout chain,
\begin{equation}
    S_\mathrm{out}=G S_\mathrm{in}.
    \label{eq:SNTJPSD}
\end{equation}

Second, we configured the switches to connect the signal input line to the readout chain again bypassing the TWPA with the thru. We measured $S_\mathrm{out}$ as a function of VNA power $P_\mathrm{VNA}$. This output power, $P_\mathrm{out}$, is a function of the gain of the readout chain, $G$, and of the attenuation $A$ on the input line, from the VNA to the TWPA input:
\begin{equation}
    P_\mathrm{out}=A G P_\mathrm{VNA}.
    \label{eq:VNAPower}
\end{equation}

\begin{figure}[H]
    \centering
    \includegraphics[width=1\linewidth]{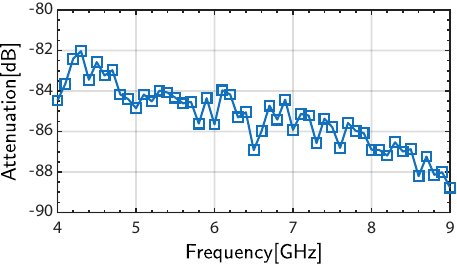}
    \caption{The calibrated input attenuation as a function of frequency from the VNA to the TWPA input reference plane was determined using the SNTJ as a calibrated power source.}
    \label{fig:inputAttenuation}
\end{figure}

Combining \eqref{eq:SNTJPSD} and \eqref{eq:VNAPower} we can calculate the input line attenuation,
\begin{equation}
    A=\frac{P_\mathrm{out}}{P_\mathrm{VNA}}\frac{S_\mathrm{in}}{S_\mathrm{out}}.
\end{equation}
Generally, this attenuation is a function of frequency, so multiple VNA power sweeps were performed at frequencies between 4 GHz and 9 GHz. The extracted attenuation is shown in Fig. \ref{fig:inputAttenuation}. Here, we have neglected the small attenuation (of order $1$\,dB \cite{malnou_low-noise_2024}) existing between the SNTJ output and the TWPA input.

\begin{figure}[H]
    \centering
    \includegraphics[width=1\linewidth]{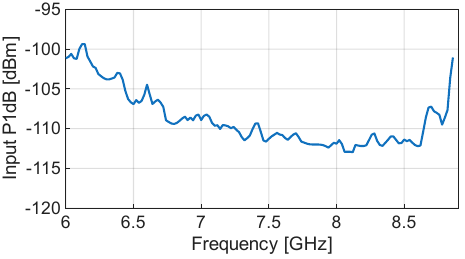}
    \caption{The $1$\,dB compression point calculation at $13$\,dB small-signal gain. The input $P_\mathrm{1dB}$ is determined using the input attenuation calibration, Fig.\ref{fig:inputAttenuation}, to convert VNA power to power at the TWPA input.}
    \label{fig:p1dB}
\end{figure}

Using this attenuation profile, we can calculate the input $1$\,dB compression power $P_\mathrm{1dB}$ by sweeping the VNA power, shown in Fig. \ref{fig:p1dB}.

% Get rid of top panel

We also used this input attenuation to approximately determine the pump power delivered to the TWPA input: assuming the cryogenic cable assemblies have nearly identical insertion loss and accounting for an additional $28$\,dB of attenuation on the signal input lines at room temperature, we estimate that the high-pass input line has $61$\,dB of attenuation at the pump frequency ($9.06$\,GHz), from the VNA to the TWPA input port. Thus, the noise-optimized pump power of $-15.8$\,dBm corresponds to $-76.8$\,dBm at the TWPA input.
\section{Diplexer}
\label{app:Diplexer}
% Add "Matched on the common port" to the diplexer info
%
%Low pass filter values:
%.param L_1 = 6.4721e-10
%.param C_2 = 5.6391e-13
%.param L_3 = 1.7565e-09
%.param C_4 = 7.1888e-13
%.param L_5 = 1.553e-09
%High pass filter values:
%.param C_1 = 6.1152e-13
%.param L_2 = 7.0186e-10
%.param C_3 = 2.2532e-13
%.param L_4 = 5.5056e-10
%.param C_5 = 2.5485e-13
%------------------------
%Capacitor Area Values
%C_1 Area ~ 2072.9626 um^2
%C_2 Area ~ 1911.5581 um^2
%C_3 Area ~ 763.8072 um^2
%C_4 Area ~ 2436.8769 um^2
%C_5 Area ~ 863.8977 um^2
%
% SINGLY TERMINATED!!!
%
% What filter, what order, singly terminated, etc.
%
% ---Single Ideas----
% Coming from the main text we already said we made a lumped element diplexer with a lowpass/highpass arrangement with the center at 8 GHz.
%
The TWPA is connected to the common ports of two identical diplexers. These diplexers are filter networks where the common port (referred to as port $1$) directs waves to one of its two other ports ($2$ and $3$) depending on the frequency $f$ of that wave relative to a crossover frequency $f_\mathrm{cx}$: ideally, for $f<f_\mathrm{cx}$ we may have $S_{21}=0$\,dB while for $f>f_\mathrm{cx}$ we may have $S_{31}=0$\,dB. We arranged each of the diplexer's filtering arms in a parallel topology \cite{matthaei_microwave_1980}, compatible with our CPW topology where the ground planes can then be left continuous and connected everywhere. The TWPA requires a low return loss at all frequencies to avoid self-resonance under pumping, and thus necessitates contiguous-band diplexers, where the end of one passband aligns with the beginning of the next passband at the crossover frequency $f_\mathrm{cx}$. In this configuration, the quality of the match will depend on the filters' orders, and will peak at $f_\mathrm{cx}$ where it is at best $-6$\,dB, because a $3$-port network cannot be simultaneously reciprocal, matched, and lossless \cite{collin_foundations_2001}. Fortunately, $f_\mathrm{cx}$ also corresponds to a low TWPA gain region. To design a contiguous-band diplexer we must use singly-terminated filters, whose components are designed for a short circuit on the common port and a load at the source port \cite{matthaei_microwave_1980}. 

\begin{figure}[]
    \centering
    \includegraphics[width=1\linewidth]{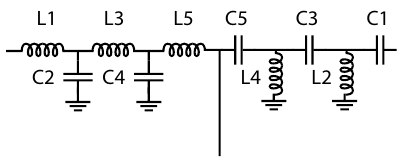}
    \caption{Component names for the lumped element diplexers.}
    \label{fig:diplexer}
\end{figure}

% The reader does not care. Don't write about what you did not do.
%A diplexer can be formed either by connecting in series filters that have a high shunt conductance in their stopbands or connecting in parallel filters that have a high series impedance in their stopbands \cite{matthaei_microwave_1980}. Our CPW circuitry lends itself best to the parallel diplexer because the ground planes can be left continuous and connected everywhere. 
%
% Diplexers can be designed where the passband of one filter ends before the passband of the next filter begins creating so-called guard bands between each channel where neither $S_{31}$ nor $S_{21}$ are near 0 dB and as a consequence of the lossless design of the filters, the input match $S_{11}$ is poor. The 4-wave mixing TWPA we designed requires a diplexer with a good input match at all frequencies, necessitating a contiguous band diplexer where the end of one passband aligns perfectly with the beginning of the next passband at the crossover frequency $f_\mathrm{cx}$. Since a 3-port network cannot be simultaneously reciprocal, matched, and lossless \cite{collin_foundations_2001,pozar_microwave_2012} the match at this crossover can be $-6$ dB at best assuming $S_{31}=S_{21}=-3$ dB as this is essentially a T-junction. In order to achieve this, we must use singly-terminated filters in the diplexer where the components are designed for a short circuit on the common port \cite{matthaei_microwave_1980}. 
%
% The two filters that make up the diplexer are 5-element Chebychev-I filters.
% Their cutoffs had to be chosen to align their 3dB point.
%
We chose singly-terminated, 5th order Chebyshev-I low-pass and high-pass filters with passband ripple $L_\mathrm{ar}= 0.1$\,dB to mitigate fabrication sensitivity, roll-off rate, and design complexity. Since the $3$\,dB frequency of Chebyshev filters does not correspond to their cutoff frequency, each filter's cutoff needed to be scaled from the $f_\mathrm{cx}=8$\,GHz crossover frequency by \cite{matthaei_microwave_1980}
\begin{equation}
    k_n = \cosh{\left(\frac{\cosh^{-1}\left(\sqrt{1/\epsilon}\right)}{n}\right)},
\end{equation}
where $\epsilon=10^{L_\mathrm{ar}/10}-1$, and $n=5$ is the filter order. Using the normalized component values $g_n$, we then calculate the actual inductor and capacitor component values. For the low-pass filter,
\begin{align}
    C_{\mathrm{n,low}} &=\frac{g_n}{Z_0 2\pi f_\mathrm{c,low}},\\
    L_{\mathrm{n,low}} &=\frac{g_n Z_0}{2\pi f_\mathrm{c,low}}
\end{align}
where $f_\mathrm{c,low}=f_\mathrm{cx}/k_5$, and where $Z_0=50$\,\ohm{} is the source impedance. Similarly for the high-pass filter,
\begin{align}
    C_{\mathrm{n,high}} &=\frac{1}{g_n Z_0 2\pi f_\mathrm{c,high}},\\
    L_{\mathrm{n,high}} &=\frac{Z_0}{g_n 2\pi f_\mathrm{c,high}},
\end{align}
where $f_\mathrm{c,high}=f_\mathrm{cx}k_5$. The prototype values and corresponding component values are shown in Tab.\,\ref{tab:diplexer} for the filter layout shown in Fig.\,\ref{fig:diplexer}.

\begin{table}[h!]
    \centering
    \begin{tabular}{c|c|c}
        $g_n$ & LPF Components & HPF Components \\
        \hline
        0.573 & $L_1=0.647$ nH & $C_1=0.612$ pF \\
        1.249 & $C_2=0.564$ pF & $L_2=0.702$ nH \\
        1.556 & $L_3=1.757$ nH & $C_3=0.225$ pF \\
        1.592 & $C_4=0.719$ pF & $L_4=0.551$ nH \\
        1.376 & $L_5=1.553$ nH & $C_5=0.255$ pF \\
    \end{tabular}
    \caption{Component values in the low-pass and high-pass diplexer arms, calculated from prototype values given in \cite{matthaei_microwave_1980}.}
    \label{tab:diplexer}
\end{table}

The layouts of these components were simulated individually, and then integrated into two electromagnetic simulations, one for each filter. These filters were then simulated together and the length between the filters was tuned to achieve the best return loss at the common port near the crossover frequency.
% Go on about how things were simulated and how the length between the filters was optimized for input match.
%
% The element values were found by applying the appropriate transformation to the prototype coefficients
%
% Once the element values were known, the required capacitor area was calculated and we used an internal lookup table to approximate each inductance. 
%
% After some tuning, here are the final simulated S-parameters
%
% ---/single ideas----
%
%Sicne we have a 4WM TWPA (and the pump is in the center of the amplification band) we want a continous-band diplexer so we lose as little of the band as possible
%
%Design Basics: An LPF in parallel with a HPF, align their 3dB points
%
%Phase adjustment: It is critical that the input impedance of each filter is high in the cutoff region at the common port (otherwise you short out the other filter)
%
%Design and Layout: We chose two 5 element filters to balance size with rolloff. The component values were chosen using a lookup table and the proper scalings for a 5th order chebychev-I filter. The cutoff frequncies were chosen sto ensure that the 3dB points aligned with one another.
%
% \bibliography{DPBib}% Produces the bibliography via BibTeX.

\begin{thebibliography}{29}%
\makeatletter
\providecommand \@ifxundefined [1]{%
 \@ifx{#1\undefined}
}%
\providecommand \@ifnum [1]{%
 \ifnum #1\expandafter \@firstoftwo
 \else \expandafter \@secondoftwo
 \fi
}%
\providecommand \@ifx [1]{%
 \ifx #1\expandafter \@firstoftwo
 \else \expandafter \@secondoftwo
 \fi
}%
\providecommand \natexlab [1]{#1}%
\providecommand \enquote  [1]{``#1''}%
\providecommand \bibnamefont  [1]{#1}%
\providecommand \bibfnamefont [1]{#1}%
\providecommand \citenamefont [1]{#1}%
\providecommand \href@noop [0]{\@secondoftwo}%
\providecommand \href [0]{\begingroup \@sanitize@url \@href}%
\providecommand \@href[1]{\@@startlink{#1}\@@href}%
\providecommand \@@href[1]{\endgroup#1\@@endlink}%
\providecommand \@sanitize@url [0]{\catcode `\\12\catcode `\$12\catcode
  `\&12\catcode `\#12\catcode `\^12\catcode `\_12\catcode `\%12\relax}%
\providecommand \@@startlink[1]{}%
\providecommand \@@endlink[0]{}%
\providecommand \url  [0]{\begingroup\@sanitize@url \@url }%
\providecommand \@url [1]{\endgroup\@href {#1}{\urlprefix }}%
\providecommand \urlprefix  [0]{URL }%
\providecommand \Eprint [0]{\href }%
\providecommand \doibase [0]{https://doi.org/}%
\def \selectlanguage #1{}%
\providecommand \bibinfo  [0]{\@secondoftwo}%
\providecommand \bibfield  [0]{\@secondoftwo}%
\providecommand \translation [1]{[#1]}%
\providecommand \BibitemOpen [0]{}%
\providecommand \bibitemStop [0]{}%
\providecommand \bibitemNoStop [0]{.\EOS\space}%
\providecommand \EOS [0]{\spacefactor3000\relax}%
\providecommand \BibitemShut  [1]{\csname bibitem#1\endcsname}%
\let\auto@bib@innerbib\@empty
%</preamble>
\bibitem [{\citenamefont {Abdo}\ \emph {et~al.}(2014)\citenamefont {Abdo},
  \citenamefont {Sliwa}, \citenamefont {Shankar}, \citenamefont {Hatridge},
  \citenamefont {Frunzio}, \citenamefont {Schoelkopf},\ and\ \citenamefont
  {Devoret}}]{Abdo2014Josephson}%
  \BibitemOpen
  \bibfield  {author} {\bibinfo {author} {\bibfnamefont {B.}~\bibnamefont
  {Abdo}}, \bibinfo {author} {\bibfnamefont {K.}~\bibnamefont {Sliwa}},
  \bibinfo {author} {\bibfnamefont {S.}~\bibnamefont {Shankar}}, \bibinfo
  {author} {\bibfnamefont {M.}~\bibnamefont {Hatridge}}, \bibinfo {author}
  {\bibfnamefont {L.}~\bibnamefont {Frunzio}}, \bibinfo {author} {\bibfnamefont
  {R.}~\bibnamefont {Schoelkopf}},\ and\ \bibinfo {author} {\bibfnamefont
  {M.}~\bibnamefont {Devoret}},\ }\bibfield  {title} {{\bibinfo {title} {Josephson {Directional} {Amplifier} for {Quantum}
  {Measurement} of {Superconducting} {Circuits}}},\ }\href
  {https://doi.org/10.1103/PhysRevLett.112.167701} {\bibfield  {journal}
  {\bibinfo  {journal} {Physical Review Letters}\ }\textbf {\bibinfo {volume}
  {112}},\ \bibinfo {pages} {167701} (\bibinfo {year} {2014})}\BibitemShut
  {NoStop}%
\bibitem [{\citenamefont {Brubaker}\ \emph {et~al.}(2017)\citenamefont
  {Brubaker}, \citenamefont {Zhong}, \citenamefont {Gurevich}, \citenamefont
  {Cahn}, \citenamefont {Lamoreaux}, \citenamefont {Simanovskaia},
  \citenamefont {Root}, \citenamefont {Lewis}, \citenamefont {Al~Kenany},
  \citenamefont {Backes}, \citenamefont {Urdinaran}, \citenamefont {Rapidis},
  \citenamefont {Shokair}, \citenamefont {Van~Bibber}, \citenamefont {Palken},
  \citenamefont {Malnou}, \citenamefont {Kindel}, \citenamefont {Anil},
  \citenamefont {Lehnert},\ and\ \citenamefont {Carosi}}]{brubaker2017first}%
  \BibitemOpen
  \bibfield  {author} {\bibinfo {author} {\bibfnamefont {B.}~\bibnamefont
  {Brubaker}}, \bibinfo {author} {\bibfnamefont {L.}~\bibnamefont {Zhong}},
  \bibinfo {author} {\bibfnamefont {Y.}~\bibnamefont {Gurevich}}, \bibinfo
  {author} {\bibfnamefont {S.}~\bibnamefont {Cahn}}, \bibinfo {author}
  {\bibfnamefont {S.}~\bibnamefont {Lamoreaux}}, \bibinfo {author}
  {\bibfnamefont {M.}~\bibnamefont {Simanovskaia}}, \bibinfo {author}
  {\bibfnamefont {J.}~\bibnamefont {Root}}, \bibinfo {author} {\bibfnamefont
  {S.}~\bibnamefont {Lewis}}, \bibinfo {author} {\bibfnamefont
  {S.}~\bibnamefont {Al~Kenany}}, \bibinfo {author} {\bibfnamefont
  {K.}~\bibnamefont {Backes}}, \bibinfo {author} {\bibfnamefont
  {I.}~\bibnamefont {Urdinaran}}, \bibinfo {author} {\bibfnamefont
  {N.}~\bibnamefont {Rapidis}}, \bibinfo {author} {\bibfnamefont
  {T.}~\bibnamefont {Shokair}}, \bibinfo {author} {\bibfnamefont
  {K.}~\bibnamefont {Van~Bibber}}, \bibinfo {author} {\bibfnamefont
  {D.}~\bibnamefont {Palken}}, \bibinfo {author} {\bibfnamefont
  {M.}~\bibnamefont {Malnou}}, \bibinfo {author} {\bibfnamefont
  {W.}~\bibnamefont {Kindel}}, \bibinfo {author} {\bibfnamefont
  {M.}~\bibnamefont {Anil}}, \bibinfo {author} {\bibfnamefont {K.}~\bibnamefont
  {Lehnert}},\ and\ \bibinfo {author} {\bibfnamefont {G.}~\bibnamefont
  {Carosi}},\ }\bibfield  {title} {{\bibinfo {title} {First
  {Results} from a {Microwave} {Cavity} {Axion} {Search} at 24 $\mu${eV}}},\
  }\href {https://doi.org/10.1103/PhysRevLett.118.061302} {\bibfield  {journal}
  {\bibinfo  {journal} {Physical Review Letters}\ }\textbf {\bibinfo {volume}
  {118}},\ \bibinfo {pages} {061302} (\bibinfo {year} {2017})}\BibitemShut
  {NoStop}%
\bibitem [{\citenamefont {Du}\ \emph {et~al.}(2018)\citenamefont {Du},
  \citenamefont {Force}, \citenamefont {Khatiwada}, \citenamefont {Lentz},
  \citenamefont {Ottens}, \citenamefont {Rosenberg}, \citenamefont {Rybka},
  \citenamefont {Carosi}, \citenamefont {Woollett}, \citenamefont {Bowring},
  \citenamefont {Chou}, \citenamefont {Sonnenschein}, \citenamefont {Wester},
  \citenamefont {Boutan}, \citenamefont {Oblath}, \citenamefont {Bradley},
  \citenamefont {Daw}, \citenamefont {Dixit}, \citenamefont {Clarke},
  \citenamefont {O’Kelley}, \citenamefont {Crisosto}, \citenamefont
  {Gleason}, \citenamefont {Jois}, \citenamefont {Sikivie}, \citenamefont
  {Stern}, \citenamefont {Sullivan}, \citenamefont {Tanner}, \citenamefont
  {Hilton},\ and\ \citenamefont {{ADMX Collaboration}}}]{du2018search}%
  \BibitemOpen
  \bibfield  {author} {\bibinfo {author} {\bibfnamefont {N.}~\bibnamefont
  {Du}}, \bibinfo {author} {\bibfnamefont {N.}~\bibnamefont {Force}}, \bibinfo
  {author} {\bibfnamefont {R.}~\bibnamefont {Khatiwada}}, \bibinfo {author}
  {\bibfnamefont {E.}~\bibnamefont {Lentz}}, \bibinfo {author} {\bibfnamefont
  {R.}~\bibnamefont {Ottens}}, \bibinfo {author} {\bibfnamefont
  {L.}~\bibnamefont {Rosenberg}}, \bibinfo {author} {\bibfnamefont
  {G.}~\bibnamefont {Rybka}}, \bibinfo {author} {\bibfnamefont
  {G.}~\bibnamefont {Carosi}}, \bibinfo {author} {\bibfnamefont
  {N.}~\bibnamefont {Woollett}}, \bibinfo {author} {\bibfnamefont
  {D.}~\bibnamefont {Bowring}}, \bibinfo {author} {\bibfnamefont
  {A.}~\bibnamefont {Chou}}, \bibinfo {author} {\bibfnamefont {A.}~\bibnamefont
  {Sonnenschein}}, \bibinfo {author} {\bibfnamefont {W.}~\bibnamefont
  {Wester}}, \bibinfo {author} {\bibfnamefont {C.}~\bibnamefont {Boutan}},
  \bibinfo {author} {\bibfnamefont {N.}~\bibnamefont {Oblath}}, \bibinfo
  {author} {\bibfnamefont {R.}~\bibnamefont {Bradley}}, \bibinfo {author}
  {\bibfnamefont {E.}~\bibnamefont {Daw}}, \bibinfo {author} {\bibfnamefont
  {A.}~\bibnamefont {Dixit}}, \bibinfo {author} {\bibfnamefont
  {J.}~\bibnamefont {Clarke}}, \bibinfo {author} {\bibfnamefont
  {S.}~\bibnamefont {O’Kelley}}, \bibinfo {author} {\bibfnamefont
  {N.}~\bibnamefont {Crisosto}}, \bibinfo {author} {\bibfnamefont
  {J.}~\bibnamefont {Gleason}}, \bibinfo {author} {\bibfnamefont
  {S.}~\bibnamefont {Jois}}, \bibinfo {author} {\bibfnamefont {P.}~\bibnamefont
  {Sikivie}}, \bibinfo {author} {\bibfnamefont {I.}~\bibnamefont {Stern}},
  \bibinfo {author} {\bibfnamefont {N.}~\bibnamefont {Sullivan}}, \bibinfo
  {author} {\bibfnamefont {D.}~\bibnamefont {Tanner}}, \bibinfo {author}
  {\bibfnamefont {G.}~\bibnamefont {Hilton}},\ and\ \bibinfo {author}
  {\bibnamefont {{ADMX Collaboration}}},\ }\bibfield  {title} {{\bibinfo {title} {Search for {Invisible} {Axion} {Dark} {Matter} with the
  {Axion} {Dark} {Matter} {Experiment}}},\ }\href
  {https://doi.org/10.1103/PhysRevLett.120.151301} {\bibfield  {journal}
  {\bibinfo  {journal} {Physical Review Letters}\ }\textbf {\bibinfo {volume}
  {120}},\ \bibinfo {pages} {151301} (\bibinfo {year} {2018})}\BibitemShut
  {NoStop}%
\bibitem [{\citenamefont {Heinsoo}\ \emph {et~al.}(2018)\citenamefont
  {Heinsoo}, \citenamefont {Andersen}, \citenamefont {Remm}, \citenamefont
  {Krinner}, \citenamefont {Walter}, \citenamefont {Salathé}, \citenamefont
  {Gasparinetti}, \citenamefont {Besse}, \citenamefont {Potočnik},
  \citenamefont {Wallraff},\ and\ \citenamefont
  {Eichler}}]{heinsoo_rapid_2018}%
  \BibitemOpen
  \bibfield  {author} {\bibinfo {author} {\bibfnamefont {J.}~\bibnamefont
  {Heinsoo}}, \bibinfo {author} {\bibfnamefont {C.~K.}\ \bibnamefont
  {Andersen}}, \bibinfo {author} {\bibfnamefont {A.}~\bibnamefont {Remm}},
  \bibinfo {author} {\bibfnamefont {S.}~\bibnamefont {Krinner}}, \bibinfo
  {author} {\bibfnamefont {T.}~\bibnamefont {Walter}}, \bibinfo {author}
  {\bibfnamefont {Y.}~\bibnamefont {Salathé}}, \bibinfo {author}
  {\bibfnamefont {S.}~\bibnamefont {Gasparinetti}}, \bibinfo {author}
  {\bibfnamefont {J.-C.}\ \bibnamefont {Besse}}, \bibinfo {author}
  {\bibfnamefont {A.}~\bibnamefont {Potočnik}}, \bibinfo {author}
  {\bibfnamefont {A.}~\bibnamefont {Wallraff}},\ and\ \bibinfo {author}
  {\bibfnamefont {C.}~\bibnamefont {Eichler}},\ }\bibfield  {title}
  {{\bibinfo {title} {Rapid {High}-fidelity {Multiplexed}
  {Readout} of {Superconducting} {Qubits}}},\ }\href
  {https://doi.org/10.1103/PhysRevApplied.10.034040} {\bibfield  {journal}
  {\bibinfo  {journal} {Physical Review Applied}\ }\textbf {\bibinfo {volume}
  {10}},\ \bibinfo {pages} {034040} (\bibinfo {year} {2018})}\BibitemShut
  {NoStop}%
\bibitem [{\citenamefont {Arute}\ \emph {et~al.}(2019)\citenamefont {Arute},
  \citenamefont {Arya}, \citenamefont {Babbush}, \citenamefont {Bacon},
  \citenamefont {Bardin}, \citenamefont {Barends}, \citenamefont {Biswas},
  \citenamefont {Boixo}, \citenamefont {Brandao}, \citenamefont {Buell},
  \citenamefont {Burkett}, \citenamefont {Chen}, \citenamefont {Chen},
  \citenamefont {Chiaro}, \citenamefont {Collins}, \citenamefont {Courtney},
  \citenamefont {Dunsworth}, \citenamefont {Farhi}, \citenamefont {Foxen},
  \citenamefont {Fowler}, \citenamefont {Gidney}, \citenamefont {Giustina},
  \citenamefont {Graff}, \citenamefont {Guerin}, \citenamefont {Habegger},
  \citenamefont {Harrigan}, \citenamefont {Hartmann}, \citenamefont {Ho},
  \citenamefont {Hoffmann}, \citenamefont {Huang}, \citenamefont {Humble},
  \citenamefont {Isakov}, \citenamefont {Jeffrey}, \citenamefont {Jiang},
  \citenamefont {Kafri}, \citenamefont {Kechedzhi}, \citenamefont {Kelly},
  \citenamefont {Klimov}, \citenamefont {Knysh}, \citenamefont {Korotkov},
  \citenamefont {Kostritsa}, \citenamefont {Landhuis}, \citenamefont
  {Lindmark}, \citenamefont {Lucero}, \citenamefont {Lyakh}, \citenamefont
  {Mandrà}, \citenamefont {McClean}, \citenamefont {McEwen}, \citenamefont
  {Megrant}, \citenamefont {Mi}, \citenamefont {Michielsen}, \citenamefont
  {Mohseni}, \citenamefont {Mutus}, \citenamefont {Naaman}, \citenamefont
  {Neeley}, \citenamefont {Neill}, \citenamefont {Niu}, \citenamefont {Ostby},
  \citenamefont {Petukhov}, \citenamefont {Platt}, \citenamefont {Quintana},
  \citenamefont {Rieffel}, \citenamefont {Roushan}, \citenamefont {Rubin},
  \citenamefont {Sank}, \citenamefont {Satzinger}, \citenamefont {Smelyanskiy},
  \citenamefont {Sung}, \citenamefont {Trevithick}, \citenamefont
  {Vainsencher}, \citenamefont {Villalonga}, \citenamefont {White},
  \citenamefont {Yao}, \citenamefont {Yeh}, \citenamefont {Zalcman},
  \citenamefont {Neven},\ and\ \citenamefont {Martinis}}]{arute2019quantum}%
  \BibitemOpen
  \bibfield  {author} {\bibinfo {author} {\bibfnamefont {F.}~\bibnamefont
  {Arute}}, \bibinfo {author} {\bibfnamefont {K.}~\bibnamefont {Arya}},
  \bibinfo {author} {\bibfnamefont {R.}~\bibnamefont {Babbush}}, \bibinfo
  {author} {\bibfnamefont {D.}~\bibnamefont {Bacon}}, \bibinfo {author}
  {\bibfnamefont {J.~C.}\ \bibnamefont {Bardin}}, \bibinfo {author}
  {\bibfnamefont {R.}~\bibnamefont {Barends}}, \bibinfo {author} {\bibfnamefont
  {R.}~\bibnamefont {Biswas}}, \bibinfo {author} {\bibfnamefont
  {S.}~\bibnamefont {Boixo}}, \bibinfo {author} {\bibfnamefont {F.~G. S.~L.}\
  \bibnamefont {Brandao}}, \bibinfo {author} {\bibfnamefont {D.~A.}\
  \bibnamefont {Buell}}, \bibinfo {author} {\bibfnamefont {B.}~\bibnamefont
  {Burkett}}, \bibinfo {author} {\bibfnamefont {Y.}~\bibnamefont {Chen}},
  \bibinfo {author} {\bibfnamefont {Z.}~\bibnamefont {Chen}}, \bibinfo {author}
  {\bibfnamefont {B.}~\bibnamefont {Chiaro}}, \bibinfo {author} {\bibfnamefont
  {R.}~\bibnamefont {Collins}}, \bibinfo {author} {\bibfnamefont
  {W.}~\bibnamefont {Courtney}}, \bibinfo {author} {\bibfnamefont
  {A.}~\bibnamefont {Dunsworth}}, \bibinfo {author} {\bibfnamefont
  {E.}~\bibnamefont {Farhi}}, \bibinfo {author} {\bibfnamefont
  {B.}~\bibnamefont {Foxen}}, \bibinfo {author} {\bibfnamefont
  {A.}~\bibnamefont {Fowler}}, \bibinfo {author} {\bibfnamefont
  {C.}~\bibnamefont {Gidney}}, \bibinfo {author} {\bibfnamefont
  {M.}~\bibnamefont {Giustina}}, \bibinfo {author} {\bibfnamefont
  {R.}~\bibnamefont {Graff}}, \bibinfo {author} {\bibfnamefont
  {K.}~\bibnamefont {Guerin}}, \bibinfo {author} {\bibfnamefont
  {S.}~\bibnamefont {Habegger}}, \bibinfo {author} {\bibfnamefont {M.~P.}\
  \bibnamefont {Harrigan}}, \bibinfo {author} {\bibfnamefont {M.~J.}\
  \bibnamefont {Hartmann}}, \bibinfo {author} {\bibfnamefont {A.}~\bibnamefont
  {Ho}}, \bibinfo {author} {\bibfnamefont {M.}~\bibnamefont {Hoffmann}},
  \bibinfo {author} {\bibfnamefont {T.}~\bibnamefont {Huang}}, \bibinfo
  {author} {\bibfnamefont {T.~S.}\ \bibnamefont {Humble}}, \bibinfo {author}
  {\bibfnamefont {S.~V.}\ \bibnamefont {Isakov}}, \bibinfo {author}
  {\bibfnamefont {E.}~\bibnamefont {Jeffrey}}, \bibinfo {author} {\bibfnamefont
  {Z.}~\bibnamefont {Jiang}}, \bibinfo {author} {\bibfnamefont
  {D.}~\bibnamefont {Kafri}}, \bibinfo {author} {\bibfnamefont
  {K.}~\bibnamefont {Kechedzhi}}, \bibinfo {author} {\bibfnamefont
  {J.}~\bibnamefont {Kelly}}, \bibinfo {author} {\bibfnamefont {P.~V.}\
  \bibnamefont {Klimov}}, \bibinfo {author} {\bibfnamefont {S.}~\bibnamefont
  {Knysh}}, \bibinfo {author} {\bibfnamefont {A.}~\bibnamefont {Korotkov}},
  \bibinfo {author} {\bibfnamefont {F.}~\bibnamefont {Kostritsa}}, \bibinfo
  {author} {\bibfnamefont {D.}~\bibnamefont {Landhuis}}, \bibinfo {author}
  {\bibfnamefont {M.}~\bibnamefont {Lindmark}}, \bibinfo {author}
  {\bibfnamefont {E.}~\bibnamefont {Lucero}}, \bibinfo {author} {\bibfnamefont
  {D.}~\bibnamefont {Lyakh}}, \bibinfo {author} {\bibfnamefont
  {S.}~\bibnamefont {Mandrà}}, \bibinfo {author} {\bibfnamefont {J.~R.}\
  \bibnamefont {McClean}}, \bibinfo {author} {\bibfnamefont {M.}~\bibnamefont
  {McEwen}}, \bibinfo {author} {\bibfnamefont {A.}~\bibnamefont {Megrant}},
  \bibinfo {author} {\bibfnamefont {X.}~\bibnamefont {Mi}}, \bibinfo {author}
  {\bibfnamefont {K.}~\bibnamefont {Michielsen}}, \bibinfo {author}
  {\bibfnamefont {M.}~\bibnamefont {Mohseni}}, \bibinfo {author} {\bibfnamefont
  {J.}~\bibnamefont {Mutus}}, \bibinfo {author} {\bibfnamefont
  {O.}~\bibnamefont {Naaman}}, \bibinfo {author} {\bibfnamefont
  {M.}~\bibnamefont {Neeley}}, \bibinfo {author} {\bibfnamefont
  {C.}~\bibnamefont {Neill}}, \bibinfo {author} {\bibfnamefont {M.~Y.}\
  \bibnamefont {Niu}}, \bibinfo {author} {\bibfnamefont {E.}~\bibnamefont
  {Ostby}}, \bibinfo {author} {\bibfnamefont {A.}~\bibnamefont {Petukhov}},
  \bibinfo {author} {\bibfnamefont {J.~C.}\ \bibnamefont {Platt}}, \bibinfo
  {author} {\bibfnamefont {C.}~\bibnamefont {Quintana}}, \bibinfo {author}
  {\bibfnamefont {E.~G.}\ \bibnamefont {Rieffel}}, \bibinfo {author}
  {\bibfnamefont {P.}~\bibnamefont {Roushan}}, \bibinfo {author} {\bibfnamefont
  {N.~C.}\ \bibnamefont {Rubin}}, \bibinfo {author} {\bibfnamefont
  {D.}~\bibnamefont {Sank}}, \bibinfo {author} {\bibfnamefont {K.~J.}\
  \bibnamefont {Satzinger}}, \bibinfo {author} {\bibfnamefont {V.}~\bibnamefont
  {Smelyanskiy}}, \bibinfo {author} {\bibfnamefont {K.~J.}\ \bibnamefont
  {Sung}}, \bibinfo {author} {\bibfnamefont {M.~D.}\ \bibnamefont
  {Trevithick}}, \bibinfo {author} {\bibfnamefont {A.}~\bibnamefont
  {Vainsencher}}, \bibinfo {author} {\bibfnamefont {B.}~\bibnamefont
  {Villalonga}}, \bibinfo {author} {\bibfnamefont {T.}~\bibnamefont {White}},
  \bibinfo {author} {\bibfnamefont {Z.~J.}\ \bibnamefont {Yao}}, \bibinfo
  {author} {\bibfnamefont {P.}~\bibnamefont {Yeh}}, \bibinfo {author}
  {\bibfnamefont {A.}~\bibnamefont {Zalcman}}, \bibinfo {author} {\bibfnamefont
  {H.}~\bibnamefont {Neven}},\ and\ \bibinfo {author} {\bibfnamefont {J.~M.}\
  \bibnamefont {Martinis}},\ }\bibfield  {title} {{\bibinfo
  {title} {Quantum supremacy using a programmable superconducting processor}},\
  }\href {https://doi.org/10.1038/s41586-019-1666-5} {\bibfield  {journal}
  {\bibinfo  {journal} {Nature}\ }\textbf {\bibinfo {volume} {574}},\ \bibinfo
  {pages} {505} (\bibinfo {year} {2019})}\BibitemShut {NoStop}%
\bibitem [{\citenamefont {Lecocq}\ \emph {et~al.}(2021)\citenamefont {Lecocq},
  \citenamefont {Ranzani}, \citenamefont {Peterson}, \citenamefont {Cicak},
  \citenamefont {Jin}, \citenamefont {Simmonds}, \citenamefont {Teufel},\ and\
  \citenamefont {Aumentado}}]{Lecocq2021efficient}%
  \BibitemOpen
  \bibfield  {author} {\bibinfo {author} {\bibfnamefont {F.}~\bibnamefont
  {Lecocq}}, \bibinfo {author} {\bibfnamefont {L.}~\bibnamefont {Ranzani}},
  \bibinfo {author} {\bibfnamefont {G.}~\bibnamefont {Peterson}}, \bibinfo
  {author} {\bibfnamefont {K.}~\bibnamefont {Cicak}}, \bibinfo {author}
  {\bibfnamefont {X.}~\bibnamefont {Jin}}, \bibinfo {author} {\bibfnamefont
  {R.}~\bibnamefont {Simmonds}}, \bibinfo {author} {\bibfnamefont
  {J.}~\bibnamefont {Teufel}},\ and\ \bibinfo {author} {\bibfnamefont
  {J.}~\bibnamefont {Aumentado}},\ }\bibfield  {title} {{\bibinfo {title} {Efficient {Qubit} {Measurement} with a {Nonreciprocal}
  {Microwave} {Amplifier}}},\ }\href
  {https://doi.org/10.1103/PhysRevLett.126.020502} {\bibfield  {journal}
  {\bibinfo  {journal} {Physical Review Letters}\ }\textbf {\bibinfo {volume}
  {126}},\ \bibinfo {pages} {020502} (\bibinfo {year} {2021})}\BibitemShut
  {NoStop}%
\bibitem [{\citenamefont {Rosenthal}\ \emph {et~al.}(2021)\citenamefont
  {Rosenthal}, \citenamefont {Schneider}, \citenamefont {Malnou}, \citenamefont
  {Zhao}, \citenamefont {Leditzky}, \citenamefont {Chapman}, \citenamefont
  {Wustmann}, \citenamefont {Ma}, \citenamefont {Palken}, \citenamefont
  {Zanner}, \citenamefont {Vale}, \citenamefont {Hilton}, \citenamefont {Gao},
  \citenamefont {Smith}, \citenamefont {Kirchmair},\ and\ \citenamefont
  {Lehnert}}]{Rosenthal2021efficient}%
  \BibitemOpen
  \bibfield  {author} {\bibinfo {author} {\bibfnamefont {E.~I.}\ \bibnamefont
  {Rosenthal}}, \bibinfo {author} {\bibfnamefont {C.~M.}\ \bibnamefont
  {Schneider}}, \bibinfo {author} {\bibfnamefont {M.}~\bibnamefont {Malnou}},
  \bibinfo {author} {\bibfnamefont {Z.}~\bibnamefont {Zhao}}, \bibinfo {author}
  {\bibfnamefont {F.}~\bibnamefont {Leditzky}}, \bibinfo {author}
  {\bibfnamefont {B.~J.}\ \bibnamefont {Chapman}}, \bibinfo {author}
  {\bibfnamefont {W.}~\bibnamefont {Wustmann}}, \bibinfo {author}
  {\bibfnamefont {X.}~\bibnamefont {Ma}}, \bibinfo {author} {\bibfnamefont
  {D.~A.}\ \bibnamefont {Palken}}, \bibinfo {author} {\bibfnamefont {M.~F.}\
  \bibnamefont {Zanner}}, \bibinfo {author} {\bibfnamefont {L.~R.}\
  \bibnamefont {Vale}}, \bibinfo {author} {\bibfnamefont {G.~C.}\ \bibnamefont
  {Hilton}}, \bibinfo {author} {\bibfnamefont {J.}~\bibnamefont {Gao}},
  \bibinfo {author} {\bibfnamefont {G.}~\bibnamefont {Smith}}, \bibinfo
  {author} {\bibfnamefont {G.}~\bibnamefont {Kirchmair}},\ and\ \bibinfo
  {author} {\bibfnamefont {K.}~\bibnamefont {Lehnert}},\ }\bibfield  {title}
  {{\bibinfo {title} {Efficient and {Low}-{Backaction}
  {Quantum} {Measurement} {Using} a {Chip}-{Scale} {Detector}}},\ }\href
  {https://doi.org/10.1103/PhysRevLett.126.090503} {\bibfield  {journal}
  {\bibinfo  {journal} {Physical Review Letters}\ }\textbf {\bibinfo {volume}
  {126}},\ \bibinfo {pages} {090503} (\bibinfo {year} {2021})}\BibitemShut
  {NoStop}%
\bibitem [{\citenamefont {Backes}\ \emph {et~al.}(2021)\citenamefont {Backes},
  \citenamefont {Palken}, \citenamefont {Kenany}, \citenamefont {Brubaker},
  \citenamefont {Cahn}, \citenamefont {Droster}, \citenamefont {Hilton},
  \citenamefont {Ghosh}, \citenamefont {Jackson}, \citenamefont {Lamoreaux},
  \citenamefont {Leder}, \citenamefont {Lehnert}, \citenamefont {Lewis},
  \citenamefont {Malnou}, \citenamefont {Maruyama}, \citenamefont {Rapidis},
  \citenamefont {Simanovskaia}, \citenamefont {Singh}, \citenamefont {Speller},
  \citenamefont {Urdinaran}, \citenamefont {Vale}, \citenamefont
  {Van~Assendelft}, \citenamefont {Van~Bibber},\ and\ \citenamefont
  {Wang}}]{Backes2021a}%
  \BibitemOpen
  \bibfield  {author} {\bibinfo {author} {\bibfnamefont {K.~M.}\ \bibnamefont
  {Backes}}, \bibinfo {author} {\bibfnamefont {D.~A.}\ \bibnamefont {Palken}},
  \bibinfo {author} {\bibfnamefont {S.~A.}\ \bibnamefont {Kenany}}, \bibinfo
  {author} {\bibfnamefont {B.~M.}\ \bibnamefont {Brubaker}}, \bibinfo {author}
  {\bibfnamefont {S.~B.}\ \bibnamefont {Cahn}}, \bibinfo {author}
  {\bibfnamefont {A.}~\bibnamefont {Droster}}, \bibinfo {author} {\bibfnamefont
  {G.~C.}\ \bibnamefont {Hilton}}, \bibinfo {author} {\bibfnamefont
  {S.}~\bibnamefont {Ghosh}}, \bibinfo {author} {\bibfnamefont
  {H.}~\bibnamefont {Jackson}}, \bibinfo {author} {\bibfnamefont {S.~K.}\
  \bibnamefont {Lamoreaux}}, \bibinfo {author} {\bibfnamefont {A.~F.}\
  \bibnamefont {Leder}}, \bibinfo {author} {\bibfnamefont {K.~W.}\ \bibnamefont
  {Lehnert}}, \bibinfo {author} {\bibfnamefont {S.~M.}\ \bibnamefont {Lewis}},
  \bibinfo {author} {\bibfnamefont {M.}~\bibnamefont {Malnou}}, \bibinfo
  {author} {\bibfnamefont {R.~H.}\ \bibnamefont {Maruyama}}, \bibinfo {author}
  {\bibfnamefont {N.~M.}\ \bibnamefont {Rapidis}}, \bibinfo {author}
  {\bibfnamefont {M.}~\bibnamefont {Simanovskaia}}, \bibinfo {author}
  {\bibfnamefont {S.}~\bibnamefont {Singh}}, \bibinfo {author} {\bibfnamefont
  {D.~H.}\ \bibnamefont {Speller}}, \bibinfo {author} {\bibfnamefont
  {I.}~\bibnamefont {Urdinaran}}, \bibinfo {author} {\bibfnamefont {L.~R.}\
  \bibnamefont {Vale}}, \bibinfo {author} {\bibfnamefont {E.~C.}\ \bibnamefont
  {Van~Assendelft}}, \bibinfo {author} {\bibfnamefont {K.}~\bibnamefont
  {Van~Bibber}},\ and\ \bibinfo {author} {\bibfnamefont {H.}~\bibnamefont
  {Wang}},\ }\bibfield  {title} {{\bibinfo {title} {A
  quantum enhanced search for dark matter axions}},\ }\href
  {https://doi.org/10.1038/s41586-021-03226-7} {\bibfield  {journal} {\bibinfo
  {journal} {Nature}\ }\textbf {\bibinfo {volume} {590}},\ \bibinfo {pages}
  {238} (\bibinfo {year} {2021})}\BibitemShut {NoStop}%
\bibitem [{\citenamefont {Krinner}\ \emph {et~al.}(2022)\citenamefont
  {Krinner}, \citenamefont {Lacroix}, \citenamefont {Remm}, \citenamefont
  {Di~Paolo}, \citenamefont {Genois}, \citenamefont {Leroux}, \citenamefont
  {Hellings}, \citenamefont {Lazar}, \citenamefont {Swiadek}, \citenamefont
  {Herrmann}, \citenamefont {Norris}, \citenamefont {Andersen}, \citenamefont
  {Müller}, \citenamefont {Blais}, \citenamefont {Eichler},\ and\
  \citenamefont {Wallraff}}]{Krinner2022realizing}%
  \BibitemOpen
  \bibfield  {author} {\bibinfo {author} {\bibfnamefont {S.}~\bibnamefont
  {Krinner}}, \bibinfo {author} {\bibfnamefont {N.}~\bibnamefont {Lacroix}},
  \bibinfo {author} {\bibfnamefont {A.}~\bibnamefont {Remm}}, \bibinfo {author}
  {\bibfnamefont {A.}~\bibnamefont {Di~Paolo}}, \bibinfo {author}
  {\bibfnamefont {E.}~\bibnamefont {Genois}}, \bibinfo {author} {\bibfnamefont
  {C.}~\bibnamefont {Leroux}}, \bibinfo {author} {\bibfnamefont
  {C.}~\bibnamefont {Hellings}}, \bibinfo {author} {\bibfnamefont
  {S.}~\bibnamefont {Lazar}}, \bibinfo {author} {\bibfnamefont
  {F.}~\bibnamefont {Swiadek}}, \bibinfo {author} {\bibfnamefont
  {J.}~\bibnamefont {Herrmann}}, \bibinfo {author} {\bibfnamefont {G.~J.}\
  \bibnamefont {Norris}}, \bibinfo {author} {\bibfnamefont {C.~K.}\
  \bibnamefont {Andersen}}, \bibinfo {author} {\bibfnamefont {M.}~\bibnamefont
  {Müller}}, \bibinfo {author} {\bibfnamefont {A.}~\bibnamefont {Blais}},
  \bibinfo {author} {\bibfnamefont {C.}~\bibnamefont {Eichler}},\ and\ \bibinfo
  {author} {\bibfnamefont {A.}~\bibnamefont {Wallraff}},\ }\bibfield  {title}
  {{\bibinfo {title} {Realizing repeated quantum error
  correction in a distance-three surface code}},\ }\href
  {https://doi.org/10.1038/s41586-022-04566-8} {\bibfield  {journal} {\bibinfo
  {journal} {Nature}\ }\textbf {\bibinfo {volume} {605}},\ \bibinfo {pages}
  {669} (\bibinfo {year} {2022})}\BibitemShut {NoStop}%
\bibitem [{\citenamefont {Malnou}\ \emph {et~al.}(2023)\citenamefont {Malnou},
  \citenamefont {Mates}, \citenamefont {Vissers}, \citenamefont {Vale},
  \citenamefont {Schmidt}, \citenamefont {Bennett}, \citenamefont {Gao},\ and\
  \citenamefont {Ullom}}]{malnou2023improved}%
  \BibitemOpen
  \bibfield  {author} {\bibinfo {author} {\bibfnamefont {M.}~\bibnamefont
  {Malnou}}, \bibinfo {author} {\bibfnamefont {J.~A.~B.}\ \bibnamefont
  {Mates}}, \bibinfo {author} {\bibfnamefont {M.~R.}\ \bibnamefont {Vissers}},
  \bibinfo {author} {\bibfnamefont {L.~R.}\ \bibnamefont {Vale}}, \bibinfo
  {author} {\bibfnamefont {D.~R.}\ \bibnamefont {Schmidt}}, \bibinfo {author}
  {\bibfnamefont {D.~A.}\ \bibnamefont {Bennett}}, \bibinfo {author}
  {\bibfnamefont {J.}~\bibnamefont {Gao}},\ and\ \bibinfo {author}
  {\bibfnamefont {J.~N.}\ \bibnamefont {Ullom}},\ }\bibfield  {title}
  {{\bibinfo {title} {Improved microwave {SQUID}
  multiplexer readout using a kinetic-inductance traveling-wave parametric
  amplifier}},\ }\href {https://doi.org/10.1063/5.0149646} {\bibfield
  {journal} {\bibinfo  {journal} {Applied Physics Letters}\ }\textbf {\bibinfo
  {volume} {122}},\ \bibinfo {pages} {214001} (\bibinfo {year}
  {2023})}\BibitemShut {NoStop}%
\bibitem [{\citenamefont {Macklin}\ \emph {et~al.}(2015)\citenamefont
  {Macklin}, \citenamefont {O’Brien}, \citenamefont {Hover}, \citenamefont
  {Schwartz}, \citenamefont {Bolkhovsky}, \citenamefont {Zhang}, \citenamefont
  {Oliver},\ and\ \citenamefont {Siddiqi}}]{macklin_nearquantum-limited_2015}%
  \BibitemOpen
  \bibfield  {author} {\bibinfo {author} {\bibfnamefont {C.}~\bibnamefont
  {Macklin}}, \bibinfo {author} {\bibfnamefont {K.}~\bibnamefont {O’Brien}},
  \bibinfo {author} {\bibfnamefont {D.}~\bibnamefont {Hover}}, \bibinfo
  {author} {\bibfnamefont {M.~E.}\ \bibnamefont {Schwartz}}, \bibinfo {author}
  {\bibfnamefont {V.}~\bibnamefont {Bolkhovsky}}, \bibinfo {author}
  {\bibfnamefont {X.}~\bibnamefont {Zhang}}, \bibinfo {author} {\bibfnamefont
  {W.~D.}\ \bibnamefont {Oliver}},\ and\ \bibinfo {author} {\bibfnamefont
  {I.}~\bibnamefont {Siddiqi}},\ }\bibfield  {title} {{\bibinfo {title} {A near–quantum-limited {Josephson} traveling-wave
  parametric amplifier}},\ }\href {https://doi.org/10.1126/science.aaa8525}
  {\bibfield  {journal} {\bibinfo  {journal} {Science}\ }\textbf {\bibinfo
  {volume} {350}},\ \bibinfo {pages} {307} (\bibinfo {year}
  {2015})}\BibitemShut {NoStop}%
\bibitem [{\citenamefont {Planat}\ \emph {et~al.}(2020)\citenamefont {Planat},
  \citenamefont {Ranadive}, \citenamefont {Dassonneville}, \citenamefont
  {Puertas~Martínez}, \citenamefont {Léger}, \citenamefont {Naud},
  \citenamefont {Buisson}, \citenamefont {Hasch-Guichard}, \citenamefont
  {Basko},\ and\ \citenamefont {Roch}}]{planat2020photonic}%
  \BibitemOpen
  \bibfield  {author} {\bibinfo {author} {\bibfnamefont {L.}~\bibnamefont
  {Planat}}, \bibinfo {author} {\bibfnamefont {A.}~\bibnamefont {Ranadive}},
  \bibinfo {author} {\bibfnamefont {R.}~\bibnamefont {Dassonneville}}, \bibinfo
  {author} {\bibfnamefont {J.}~\bibnamefont {Puertas~Martínez}}, \bibinfo
  {author} {\bibfnamefont {S.}~\bibnamefont {Léger}}, \bibinfo {author}
  {\bibfnamefont {C.}~\bibnamefont {Naud}}, \bibinfo {author} {\bibfnamefont
  {O.}~\bibnamefont {Buisson}}, \bibinfo {author} {\bibfnamefont
  {W.}~\bibnamefont {Hasch-Guichard}}, \bibinfo {author} {\bibfnamefont
  {D.~M.}\ \bibnamefont {Basko}},\ and\ \bibinfo {author} {\bibfnamefont
  {N.}~\bibnamefont {Roch}},\ }\bibfield  {title} {{\bibinfo {title} {Photonic-{Crystal} {Josephson} {Traveling}-{Wave}
  {Parametric} {Amplifier}}},\ }\href
  {https://doi.org/10.1103/PhysRevX.10.021021} {\bibfield  {journal} {\bibinfo
  {journal} {Physical Review X}\ }\textbf {\bibinfo {volume} {10}},\ \bibinfo
  {pages} {021021} (\bibinfo {year} {2020})}\BibitemShut {NoStop}%
\bibitem [{\citenamefont {Malnou}\ \emph {et~al.}(2021)\citenamefont {Malnou},
  \citenamefont {Vissers}, \citenamefont {Wheeler}, \citenamefont {Aumentado},
  \citenamefont {Hubmayr}, \citenamefont {Ullom},\ and\ \citenamefont
  {Gao}}]{malnou_three-wave_2021}%
  \BibitemOpen
  \bibfield  {author} {\bibinfo {author} {\bibfnamefont {M.}~\bibnamefont
  {Malnou}}, \bibinfo {author} {\bibfnamefont {M.}~\bibnamefont {Vissers}},
  \bibinfo {author} {\bibfnamefont {J.}~\bibnamefont {Wheeler}}, \bibinfo
  {author} {\bibfnamefont {J.}~\bibnamefont {Aumentado}}, \bibinfo {author}
  {\bibfnamefont {J.}~\bibnamefont {Hubmayr}}, \bibinfo {author} {\bibfnamefont
  {J.}~\bibnamefont {Ullom}},\ and\ \bibinfo {author} {\bibfnamefont
  {J.}~\bibnamefont {Gao}},\ }\bibfield  {title} {{\bibinfo
  {title} {Three-{Wave} {Mixing} {Kinetic} {Inductance} {Traveling}-{Wave}
  {Amplifier} with {Near}-{Quantum}-{Limited} {Noise} {Performance}}},\ }\href
  {https://doi.org/10.1103/PRXQuantum.2.010302} {\bibfield  {journal} {\bibinfo
   {journal} {PRX Quantum}\ }\textbf {\bibinfo {volume} {2}},\ \bibinfo {pages}
  {010302} (\bibinfo {year} {2021})}\BibitemShut {NoStop}%
\bibitem [{\citenamefont {Malnou}\ \emph {et~al.}(2024)\citenamefont {Malnou},
  \citenamefont {Larson}, \citenamefont {Teufel}, \citenamefont {Lecocq},\ and\
  \citenamefont {Aumentado}}]{malnou_low-noise_2024}%
  \BibitemOpen
  \bibfield  {author} {\bibinfo {author} {\bibfnamefont {M.}~\bibnamefont
  {Malnou}}, \bibinfo {author} {\bibfnamefont {T.~F.~Q.}\ \bibnamefont
  {Larson}}, \bibinfo {author} {\bibfnamefont {J.~D.}\ \bibnamefont {Teufel}},
  \bibinfo {author} {\bibfnamefont {F.}~\bibnamefont {Lecocq}},\ and\ \bibinfo
  {author} {\bibfnamefont {J.}~\bibnamefont {Aumentado}},\ }\bibfield  {title}
  {{\bibinfo {title} {Low-noise cryogenic microwave
  amplifier characterization with a calibrated noise source}},\ }\href
  {https://doi.org/10.1063/5.0193591} {\bibfield  {journal} {\bibinfo
  {journal} {Review of Scientific Instruments}\ }\textbf {\bibinfo {volume}
  {95}},\ \bibinfo {pages} {034703} (\bibinfo {year} {2024})}\BibitemShut
  {NoStop}%
\bibitem [{\citenamefont {Malnou}\ \emph {et~al.}(2025)\citenamefont {Malnou},
  \citenamefont {Miller}, \citenamefont {Estrada}, \citenamefont {Genter},
  \citenamefont {Cicak}, \citenamefont {Teufel}, \citenamefont {Aumentado},\
  and\ \citenamefont {Lecocq}}]{malnou_travelling-wave_2025}%
  \BibitemOpen
  \bibfield  {author} {\bibinfo {author} {\bibfnamefont {M.}~\bibnamefont
  {Malnou}}, \bibinfo {author} {\bibfnamefont {B.~T.}\ \bibnamefont {Miller}},
  \bibinfo {author} {\bibfnamefont {J.~A.}\ \bibnamefont {Estrada}}, \bibinfo
  {author} {\bibfnamefont {K.}~\bibnamefont {Genter}}, \bibinfo {author}
  {\bibfnamefont {K.}~\bibnamefont {Cicak}}, \bibinfo {author} {\bibfnamefont
  {J.~D.}\ \bibnamefont {Teufel}}, \bibinfo {author} {\bibfnamefont
  {J.}~\bibnamefont {Aumentado}},\ and\ \bibinfo {author} {\bibfnamefont
  {F.}~\bibnamefont {Lecocq}},\ }\bibfield  {title} {{\bibinfo {title} {A travelling-wave parametric amplifier and
  converter}},\ }\href {https://doi.org/10.1038/s41928-025-01445-8} {\bibfield
  {journal} {\bibinfo  {journal} {Nature Electronics}\ }\textbf {\bibinfo
  {volume} {8}},\ \bibinfo {pages} {1082} (\bibinfo {year} {2025})}\BibitemShut
  {NoStop}%
\bibitem [{\citenamefont {Ranadive}\ \emph {et~al.}(2025)\citenamefont
  {Ranadive}, \citenamefont {Fazliji}, \citenamefont {Le~Gal}, \citenamefont
  {Cappelli}, \citenamefont {Butseraen}, \citenamefont {Bonet}, \citenamefont
  {Eyraud}, \citenamefont {Böhling}, \citenamefont {Planat}, \citenamefont
  {Metelmann},\ and\ \citenamefont {Roch}}]{Ranadive2025a}%
  \BibitemOpen
  \bibfield  {author} {\bibinfo {author} {\bibfnamefont {A.}~\bibnamefont
  {Ranadive}}, \bibinfo {author} {\bibfnamefont {B.}~\bibnamefont {Fazliji}},
  \bibinfo {author} {\bibfnamefont {G.}~\bibnamefont {Le~Gal}}, \bibinfo
  {author} {\bibfnamefont {G.}~\bibnamefont {Cappelli}}, \bibinfo {author}
  {\bibfnamefont {G.}~\bibnamefont {Butseraen}}, \bibinfo {author}
  {\bibfnamefont {E.}~\bibnamefont {Bonet}}, \bibinfo {author} {\bibfnamefont
  {E.}~\bibnamefont {Eyraud}}, \bibinfo {author} {\bibfnamefont
  {S.}~\bibnamefont {Böhling}}, \bibinfo {author} {\bibfnamefont
  {L.}~\bibnamefont {Planat}}, \bibinfo {author} {\bibfnamefont
  {A.}~\bibnamefont {Metelmann}},\ and\ \bibinfo {author} {\bibfnamefont
  {N.}~\bibnamefont {Roch}},\ }\bibfield  {title} {{\bibinfo {title} {A travelling-wave parametric amplifier isolator}},\
  }\href {https://doi.org/10.1038/s41928-025-01489-w} {\bibfield  {journal}
  {\bibinfo  {journal} {Nature Electronics}\ }\textbf {\bibinfo {volume} {8}},\
  \bibinfo {pages} {1089} (\bibinfo {year} {2025})}\BibitemShut {NoStop}%
\bibitem [{\citenamefont {Spietz}\ \emph {et~al.}(2003)\citenamefont {Spietz},
  \citenamefont {Lehnert}, \citenamefont {Siddiqi},\ and\ \citenamefont
  {Schoelkopf}}]{Spietz2003primary}%
  \BibitemOpen
  \bibfield  {author} {\bibinfo {author} {\bibfnamefont {L.}~\bibnamefont
  {Spietz}}, \bibinfo {author} {\bibfnamefont {K.~W.}\ \bibnamefont {Lehnert}},
  \bibinfo {author} {\bibfnamefont {I.}~\bibnamefont {Siddiqi}},\ and\ \bibinfo
  {author} {\bibfnamefont {R.~J.}\ \bibnamefont {Schoelkopf}},\ }\bibfield
  {title} {{\bibinfo {title} {Primary {Electronic}
  {Thermometry} {Using} the {Shot} {Noise} of a {Tunnel} {Junction}}},\ }\href
  {https://doi.org/10.1126/science.1084647} {\bibfield  {journal} {\bibinfo
  {journal} {Science}\ }\textbf {\bibinfo {volume} {300}},\ \bibinfo {pages}
  {1929} (\bibinfo {year} {2003})}\BibitemShut {NoStop}%
\bibitem [{\citenamefont {Spietz}\ \emph {et~al.}(2006)\citenamefont {Spietz},
  \citenamefont {Schoelkopf},\ and\ \citenamefont {Pari}}]{spietz2006shot}%
  \BibitemOpen
  \bibfield  {author} {\bibinfo {author} {\bibfnamefont {L.}~\bibnamefont
  {Spietz}}, \bibinfo {author} {\bibfnamefont {R.~J.}\ \bibnamefont
  {Schoelkopf}},\ and\ \bibinfo {author} {\bibfnamefont {P.}~\bibnamefont
  {Pari}},\ }\bibfield  {title} {{\bibinfo {title} {Shot
  noise thermometry down to {10mK}}},\ }\href
  {https://doi.org/10.1063/1.2382736} {\bibfield  {journal} {\bibinfo
  {journal} {Applied Physics Letters}\ }\textbf {\bibinfo {volume} {89}},\
  \bibinfo {pages} {183123} (\bibinfo {year} {2006})}\BibitemShut {NoStop}%
\bibitem [{\citenamefont {{Su-Wei Chang}}\ \emph {et~al.}(2016)\citenamefont
  {{Su-Wei Chang}}, \citenamefont {Aumentado}, \citenamefont {{Wei-Ting
  Wong}},\ and\ \citenamefont {Bardin}}]{chang2016noise}%
  \BibitemOpen
  \bibfield  {author} {\bibinfo {author} {\bibnamefont {{Su-Wei Chang}}},
  \bibinfo {author} {\bibfnamefont {J.}~\bibnamefont {Aumentado}}, \bibinfo
  {author} {\bibnamefont {{Wei-Ting Wong}}},\ and\ \bibinfo {author}
  {\bibfnamefont {J.~C.}\ \bibnamefont {Bardin}},\ }\bibfield  {title}
  {\bibinfo {title} {Noise measurement of cryogenic low noise amplifiers using
  a tunnel-junction shot-noise source},\ }in\ \href
  {https://doi.org/10.1109/MWSYM.2016.7538226} {\emph {\bibinfo {booktitle}
  {2016 {IEEE} {MTT}-{S} {International} {Microwave} {Symposium} ({IMS})}}}\
  (\bibinfo  {publisher} {IEEE},\ \bibinfo {address} {San Francisco, CA},\
  \bibinfo {year} {2016})\ pp.\ \bibinfo {pages} {1--4}\BibitemShut {NoStop}%
\bibitem [{\citenamefont {Yaakobi}\ \emph {et~al.}(2013)\citenamefont
  {Yaakobi}, \citenamefont {Friedland}, \citenamefont {Macklin},\ and\
  \citenamefont {Siddiqi}}]{yaakobi_parametric_2013}%
  \BibitemOpen
  \bibfield  {author} {\bibinfo {author} {\bibfnamefont {O.}~\bibnamefont
  {Yaakobi}}, \bibinfo {author} {\bibfnamefont {L.}~\bibnamefont {Friedland}},
  \bibinfo {author} {\bibfnamefont {C.}~\bibnamefont {Macklin}},\ and\ \bibinfo
  {author} {\bibfnamefont {I.}~\bibnamefont {Siddiqi}},\ }\bibfield  {title}
  {{\bibinfo {title} {Parametric amplification in
  {Josephson} junction embedded transmission lines}},\ }\href
  {https://doi.org/10.1103/PhysRevB.87.144301} {\bibfield  {journal} {\bibinfo
  {journal} {Physical Review B}\ }\textbf {\bibinfo {volume} {87}},\ \bibinfo
  {pages} {144301} (\bibinfo {year} {2013})}\BibitemShut {NoStop}%
\bibitem [{\citenamefont {O’Brien}\ \emph {et~al.}(2014)\citenamefont
  {O’Brien}, \citenamefont {Macklin}, \citenamefont {Siddiqi},\ and\
  \citenamefont {Zhang}}]{obrien_resonant_2014}%
  \BibitemOpen
  \bibfield  {author} {\bibinfo {author} {\bibfnamefont {K.}~\bibnamefont
  {O’Brien}}, \bibinfo {author} {\bibfnamefont {C.}~\bibnamefont {Macklin}},
  \bibinfo {author} {\bibfnamefont {I.}~\bibnamefont {Siddiqi}},\ and\ \bibinfo
  {author} {\bibfnamefont {X.}~\bibnamefont {Zhang}},\ }\bibfield  {title}
  {{\bibinfo {title} {Resonant {Phase} {Matching} of
  {Josephson} {Junction} {Traveling} {Wave} {Parametric} {Amplifiers}}},\
  }\href {https://doi.org/10.1103/PhysRevLett.113.157001} {\bibfield  {journal}
  {\bibinfo  {journal} {Physical Review Letters}\ }\textbf {\bibinfo {volume}
  {113}},\ \bibinfo {pages} {157001} (\bibinfo {year} {2014})}\BibitemShut
  {NoStop}%
\bibitem [{\citenamefont {Lecocq}\ \emph {et~al.}(2017)\citenamefont {Lecocq},
  \citenamefont {Ranzani}, \citenamefont {Peterson}, \citenamefont {Cicak},
  \citenamefont {Simmonds}, \citenamefont {Teufel},\ and\ \citenamefont
  {Aumentado}}]{lecocq2017nonreciprocal}%
  \BibitemOpen
  \bibfield  {author} {\bibinfo {author} {\bibfnamefont {F.}~\bibnamefont
  {Lecocq}}, \bibinfo {author} {\bibfnamefont {L.}~\bibnamefont {Ranzani}},
  \bibinfo {author} {\bibfnamefont {G.}~\bibnamefont {Peterson}}, \bibinfo
  {author} {\bibfnamefont {K.}~\bibnamefont {Cicak}}, \bibinfo {author}
  {\bibfnamefont {R.}~\bibnamefont {Simmonds}}, \bibinfo {author}
  {\bibfnamefont {J.}~\bibnamefont {Teufel}},\ and\ \bibinfo {author}
  {\bibfnamefont {J.}~\bibnamefont {Aumentado}},\ }\bibfield  {title}
  {{\bibinfo {title} {Nonreciprocal {Microwave} {Signal}
  {Processing} with a {Field}-{Programmable} {Josephson} {Amplifier}}},\ }\href
  {https://doi.org/10.1103/PhysRevApplied.7.024028} {\bibfield  {journal}
  {\bibinfo  {journal} {Physical Review Applied}\ }\textbf {\bibinfo {volume}
  {7}},\ \bibinfo {pages} {024028} (\bibinfo {year} {2017})}\BibitemShut
  {NoStop}%
\bibitem [{\citenamefont {Peng}\ \emph {et~al.}(2022)\citenamefont {Peng},
  \citenamefont {Naghiloo}, \citenamefont {Wang}, \citenamefont {Cunningham},
  \citenamefont {Ye},\ and\ \citenamefont
  {O’Brien}}]{peng_floquet-mode_2022}%
  \BibitemOpen
  \bibfield  {author} {\bibinfo {author} {\bibfnamefont {K.}~\bibnamefont
  {Peng}}, \bibinfo {author} {\bibfnamefont {M.}~\bibnamefont {Naghiloo}},
  \bibinfo {author} {\bibfnamefont {J.}~\bibnamefont {Wang}}, \bibinfo {author}
  {\bibfnamefont {G.~D.}\ \bibnamefont {Cunningham}}, \bibinfo {author}
  {\bibfnamefont {Y.}~\bibnamefont {Ye}},\ and\ \bibinfo {author}
  {\bibfnamefont {K.~P.}\ \bibnamefont {O’Brien}},\ }\bibfield  {title}
  {{\bibinfo {title} {Floquet-{Mode} {Traveling}-{Wave}
  {Parametric} {Amplifiers}}},\ }\href
  {https://doi.org/10.1103/PRXQuantum.3.020306} {\bibfield  {journal} {\bibinfo
   {journal} {PRX Quantum}\ }\textbf {\bibinfo {volume} {3}},\ \bibinfo {pages}
  {020306} (\bibinfo {year} {2022})}\BibitemShut {NoStop}%
\bibitem [{\citenamefont {Wang}\ \emph {et~al.}(2025)\citenamefont {Wang},
  \citenamefont {Peng}, \citenamefont {Knecht}, \citenamefont {Cunningham},
  \citenamefont {Lombo}, \citenamefont {Yen}, \citenamefont {Zaidenberg},
  \citenamefont {Gingras}, \citenamefont {Niedzielski}, \citenamefont
  {Stickler}, \citenamefont {Sliwa}, \citenamefont {Serniak}, \citenamefont
  {Schwartz}, \citenamefont {Oliver},\ and\ \citenamefont
  {O'Brien}}]{wang2025high}%
  \BibitemOpen
  \bibfield  {author} {\bibinfo {author} {\bibfnamefont {J.}~\bibnamefont
  {Wang}}, \bibinfo {author} {\bibfnamefont {K.}~\bibnamefont {Peng}}, \bibinfo
  {author} {\bibfnamefont {J.~M.}\ \bibnamefont {Knecht}}, \bibinfo {author}
  {\bibfnamefont {G.~D.}\ \bibnamefont {Cunningham}}, \bibinfo {author}
  {\bibfnamefont {A.~E.}\ \bibnamefont {Lombo}}, \bibinfo {author}
  {\bibfnamefont {A.}~\bibnamefont {Yen}}, \bibinfo {author} {\bibfnamefont
  {D.~A.}\ \bibnamefont {Zaidenberg}}, \bibinfo {author} {\bibfnamefont
  {M.}~\bibnamefont {Gingras}}, \bibinfo {author} {\bibfnamefont {B.~M.}\
  \bibnamefont {Niedzielski}}, \bibinfo {author} {\bibfnamefont
  {H.}~\bibnamefont {Stickler}}, \bibinfo {author} {\bibfnamefont
  {K.}~\bibnamefont {Sliwa}}, \bibinfo {author} {\bibfnamefont
  {K.}~\bibnamefont {Serniak}}, \bibinfo {author} {\bibfnamefont {M.~E.}\
  \bibnamefont {Schwartz}}, \bibinfo {author} {\bibfnamefont {W.~D.}\
  \bibnamefont {Oliver}},\ and\ \bibinfo {author} {\bibfnamefont {K.~P.}\
  \bibnamefont {O'Brien}},\ }\href {https://arxiv.org/abs/2503.11812} {\bibinfo
  {title} {High-efficiency, low-loss floquet-mode traveling wave parametric
  amplifier}} (\bibinfo {year} {2025}),\ \Eprint
  {https://arxiv.org/abs/2503.11812} {arXiv:2503.11812 [quant-ph]} \BibitemShut
  {NoStop}%
\bibitem [{\citenamefont {{Whiteley Research Inc.}}(2025)}]{wrspice}%
  \BibitemOpen
  \bibfield  {author} {\bibinfo {author} {\bibnamefont {{Whiteley Research
  Inc.}}},\ }\href@noop {} {\bibinfo {title} {{WRspice Circuit Simulation
  Program}}},\ \bibinfo {howpublished} {\url{http://www.wrcad.com}} (\bibinfo
  {year} {2025}),\ \bibinfo {note} {sunnyvale, CA}\BibitemShut {NoStop}%
\bibitem [{\citenamefont {Gaydamachenko}\ \emph {et~al.}(2025)\citenamefont
  {Gaydamachenko}, \citenamefont {Kissling},\ and\ \citenamefont
  {Grünhaupt}}]{gaydamachenko_rf-squid-based_2025}%
  \BibitemOpen
  \bibfield  {author} {\bibinfo {author} {\bibfnamefont {V.}~\bibnamefont
  {Gaydamachenko}}, \bibinfo {author} {\bibfnamefont {C.}~\bibnamefont
  {Kissling}},\ and\ \bibinfo {author} {\bibfnamefont {L.}~\bibnamefont
  {Grünhaupt}},\ }\bibfield  {title} {{\bibinfo {title}
  {rf-{SQUID}-based traveling-wave parametric amplifier with input saturation
  power of $-$84 {dBm} across more than one octave in bandwidth}},\ }\href
  {https://doi.org/10.1103/1qk4-fzkq} {\bibfield  {journal} {\bibinfo
  {journal} {Physical Review Applied}\ }\textbf {\bibinfo {volume} {23}},\
  \bibinfo {pages} {064053} (\bibinfo {year} {2025})}\BibitemShut {NoStop}%
\bibitem [{\citenamefont {Chang}\ \emph {et~al.}(2025)\citenamefont {Chang},
  \citenamefont {Van~Loo}, \citenamefont {Hung}, \citenamefont {Zhou},
  \citenamefont {Gnandt}, \citenamefont {Tamate},\ and\ \citenamefont
  {Nakamura}}]{chang_josephson_2025}%
  \BibitemOpen
  \bibfield  {author} {\bibinfo {author} {\bibfnamefont {C.~S.}\ \bibnamefont
  {Chang}}, \bibinfo {author} {\bibfnamefont {A.~F.}\ \bibnamefont {Van~Loo}},
  \bibinfo {author} {\bibfnamefont {C.-C.}\ \bibnamefont {Hung}}, \bibinfo
  {author} {\bibfnamefont {Y.}~\bibnamefont {Zhou}}, \bibinfo {author}
  {\bibfnamefont {C.}~\bibnamefont {Gnandt}}, \bibinfo {author} {\bibfnamefont
  {S.}~\bibnamefont {Tamate}},\ and\ \bibinfo {author} {\bibfnamefont
  {Y.}~\bibnamefont {Nakamura}},\ }\bibfield  {title} {{\bibinfo {title} {Josephson traveling-wave parametric amplifier based on
  a low-intrinsic-loss lumped-element coplanar waveguide}},\ }\href
  {https://doi.org/10.1103/qhl6-cz2z} {\bibfield  {journal} {\bibinfo
  {journal} {Physical Review Applied}\ }\textbf {\bibinfo {volume} {24}},\
  \bibinfo {pages} {044081} (\bibinfo {year} {2025})}\BibitemShut {NoStop}%
\bibitem [{\citenamefont {Matthaei}\ \emph {et~al.}(1980)\citenamefont
  {Matthaei}, \citenamefont {Young},\ and\ \citenamefont
  {Jones}}]{matthaei_microwave_1980}%
  \BibitemOpen
  \bibfield  {author} {\bibinfo {author} {\bibfnamefont {G.~L.}\ \bibnamefont
  {Matthaei}}, \bibinfo {author} {\bibfnamefont {L.}~\bibnamefont {Young}},\
  and\ \bibinfo {author} {\bibfnamefont {E.~M.~T.}\ \bibnamefont {Jones}},\
  }\href@noop {} {{\emph {\bibinfo {title} {Microwave
  filters, impedance-matching networks, and coupling structures}}}},\ \bibinfo
  {edition} {facsim. ed}\ ed.,\ The {Artech} house microwave library\ (\bibinfo
   {publisher} {Artech},\ \bibinfo {address} {Dedham (Mass.)},\ \bibinfo {year}
  {1980})\BibitemShut {NoStop}%
\bibitem [{\citenamefont {Collin}(2001)}]{collin_foundations_2001}%
  \BibitemOpen
  \bibfield  {author} {\bibinfo {author} {\bibfnamefont {R.~E.}\ \bibnamefont
  {Collin}},\ }\href {https://doi.org/10.1109/9780470544662} {{\emph {\bibinfo {title} {Foundations for microwave engineering}}}},\
  \bibinfo {edition} {second edition}\ ed.,\ {IEEE} {Press} series on
  electromagnetic wave theory\ (\bibinfo  {publisher} {IEEE Press,
  Wiley-Interscience},\ \bibinfo {address} {New York Hoboken, New Jersey},\
  \bibinfo {year} {2001})\BibitemShut {NoStop}%
\label{LastBibItem}%
\end{thebibliography}


%
\end{document}